\documentclass[11pt,journal,draftcls,onecolumn,peerreviewca]{IEEEtran}
\usepackage[T1]{fontenc}
\usepackage{algorithm}
\usepackage{algorithmic}
\usepackage{amsmath}
\usepackage{amsthm}
\usepackage{amssymb}
\usepackage{bbm}
\usepackage{color}
\usepackage{diagbox}
\usepackage{esint}
\usepackage{float}
\usepackage{graphicx}
\usepackage{mathrsfs}
\usepackage{microtype}
\usepackage{multicol}
\usepackage{stackrel}
\usepackage{stfloats}
\usepackage{stmaryrd}
\usepackage{subfigure}
\usepackage{verbatim}

\newtheorem{proposition}{Proposition}

\begin{document}
\title{Line Spectral Estimation via Unlimited Sampling}
\author{Qi Zhang, Jiang Zhu, Fengzhong Qu, and De Wen Soh
\thanks{Qi Zhang and De Wen Soh are with the Information Systems Technology and Design, Singapore University of Technology and Design, 487372, Singapore (email: \{qi\_zhang@mymail.sutd.edu.sg, dewen\_soh@sutd.edu.sg\}).

Jiang Zhu and Fengzhong Qu are with the Ocean College, Zhejiang University, No.1 Zheda Road, Zhoushan, 316021, China (email: \{jiangzhu16, jimqufz\}@zju.edu.cn).}}
\maketitle
\begin{abstract}
  Frequency Modulated Continuous Wave (FMCW) radar has been widely applied in automotive anti-collision systems, automatic cruise control, and indoor monitoring. However, conventional analog-to-digital converters (ADCs) can suffer from significant information loss when strong and weak targets coexist in ranging applications. To address this issue, the Unlimited Sampling (US) strategy was proposed, which applies a modulo operator prior to sampling. In this paper, we investigate the range estimation problem using FMCW radar in the context of US, which can be formulated as a one-dimensional line spectral estimation (LSE) via US. By exploiting the oversampling property and proving that the leakage onto a certain frequency band can be controlled, we establish an integer optimization problem in the Fourier and first-order difference domain. We then propose a dynamic programming (DP) based algorithm followed by the orthogonal matching pursuit (OMP) method to solve it. In addition, a two-stage US LSE (USLSE) is proposed, where the line spectral signal is first recovered by iteratively executing DP and OMP, and then the parameters are estimated by applying a state-of-the-art LSE algorithm. Substantial numerical simulations and real experiments demonstrate that the proposed algorithm, USLSE, outperforms existing algorithms. 
\end{abstract}
\section{Introduction}
Frequency Modulated Continuous Wave (FMCW) radar has been widely applied in automotive anti-collision systems, automatic cruise control, and indoor monitoring due to its high ranging precision and resolution, outstanding interference resistance, and simple structure and compact size \cite{stove1992linear, bar2002doa, geroleo2012detection, jin2020one}. In ranging applications, the presence of the inverse-square law, where the intensity of a target is inversely proportional to the square of the distance from the target to the source, and variations in the reflective properties of different materials often result in scenarios where strong and weak targets coexist \cite{zhang2019range, feuillen2022unlimited}. Such scenarios include detecting humans near trucks, humans across a broad range, and indoor monitoring amid significant clutter \cite{engels2017advances, vandersmissen2018indoor}.

For the conventional analog-to-digital converter (ADC) used in FMCW radar systems, the information loss is inevitable in the scenario where strong and weak targets coexist due to the dilemma below \cite{block2006performance}: On the one hand, due to the exist of strong targets, the amplitude of the signal may exceed the dynamic range of the ADC which will lead to saturation or clipping, resulting in the loss of information from the most significant bits (MSBs). On the other hand, if we adjust the dynamic range of the ADC to avoid clipping, the weak signal may be submerged in quantization noise, causing the loss of information from the least significant bits (LSBs). To address this problem, self-reset ADCs are implemented in which a modulo operation is applied before sampling \cite{park2007wide, yuan2009activity, sasagawa2015implantable, krishna2019unlimited}. In these architectures, side information like the amount of folding for each sample or the folding sign is stored for perfect reconstruction, which requires complex electronic circuitry. 

Recently, signal recovery from modulo samples only has been studied in depth. The theory that there is a one-to-one mapping between modulo samples and unfolded measurements has been established when the sampling rate is slightly above the critical Nyquist rate for bandlimited and finite energy signals \cite{bhandari2019identifiability,romanov2019above}. In addition, if the number of samples per interval is equal to or exceeds $2L + 1$ and $2L + 1$ is a prime number, the $L$-order periodic bandlimited signal can be uniquely identiﬁable up to a constant factor from the modulo samples \cite{mulleti2022modulo}. Furthermore, from the perspective of compressed sensing, researchers have derived the minimum number of modulo measurements required to achieve the recovery guarantee for sparse vectors \cite{prasanna2020identifiability}. 

Algorithms for reconstruction can mainly be divided into two categories, one is based on the time domain processing, and the other is based on the frequency domain processing. In the time domain, Bhandari et al. proposed an unlimited sampling algorithm (USAlg) for bandlimited signals, which is based on the higher-order differences operator, and provided a recovery guarantee when the oversampling rate exceeds $2\pi e$ times the Nyquist rate \cite{bhandari2020unlimited}. USAlg has been extended to address different signal models and problems, such as the mixture of sinusoids, finite-rate-of-innovation signals, multidimensional signals, DOA estimation, computed tomography, graph signals
\cite{bhandari2018unlimited1,bhandari2018unlimited2,bouis2021multidimensional,fernandez2021doa,bhandari2021modulo,fernandez2022computational,beckmann2020hdr,bhandari2020hdr}, etc. However, due to higher-order differences, USAlg is sensitive to noise, especially when the noise is added to the original signal. Ordentlich et al. proposed another method in the time domain named integer-forcing decoding to unwrap the random Gaussian distributed vectors, where the covariance matrix is known \cite{ordentlich2016integer, ordentlich2018modulo}. The blind version is also proposed where an empirical method is used to estimate the covariance matrix \cite{romanov2021blind}. In addition, the linear prediction (LP) approach is introduced for stationary stochastic Gaussian processes in both the temporal and spatio-temporal cases, and approaches the limits given by rate-distortion theory \cite{ordentlich2018modulo}. The LP approach is also applied for bandlimited signals and the recovery guarantee is provided when the oversampling factor is greater than $1$ \cite{romanov2019above}. The LP method requires that the initial few samples of the signal are always within the dynamic range, and the autocorrelation function is known. Furthermore, a blind version of the LP method is proposed which incorporates both an automatic adjustment at the modulo level and a fully adaptable modulo unwrapping mechanism \cite{weiss2022blind2,weiss2022blind}. For the recovery approach based on the Fourier domain, Bhandari et al. proposed an algorithm for recovering the periodic bandlimited signals by solving a spectral estimation problem with Prony's method when the number of folds is known, and this method can be used to handle the non-ideal foldings in practical hardware \cite{bhandari2021unlimited}. Another kind of method in the Fourier domain uses the projected gradient descent method to estimate the residual \cite{azar2022residual,azar2022robust}. Besides, to recover the clean original signal from noisy modulo samples, two-stage algorithms are proposed which perform denoising on modulo samples before recovering \cite{cucuringu2020provably, fanuel2022denoising}.

Hardware prototype for unlimited sampling is also designed, which is first introduced in \cite{bhandari2021unlimited} followed by \cite{mulleti2023hardware}. To deal with hysteresis and folding transients in practical hardware, a thresholding approach with a recovery guarantee is proposed which can also be applied to the case when the non-idealities are not present, and a low sampling rate version is also provided \cite{florescu2022surprising}. In addition, thresholding with general filters and average sampling are utilized to deal with non-idealities \cite{florescu2022unlimited1, florescu2022unlimited2}. Considering the practical hardware, three signal folding architectures are summarized in \cite{shtendel2023unlimited}, and the reconstruction algorithms are proposed for bandpass signals.

\begin{figure*}[t!]
		\centering
		\includegraphics[width=0.85\textwidth]{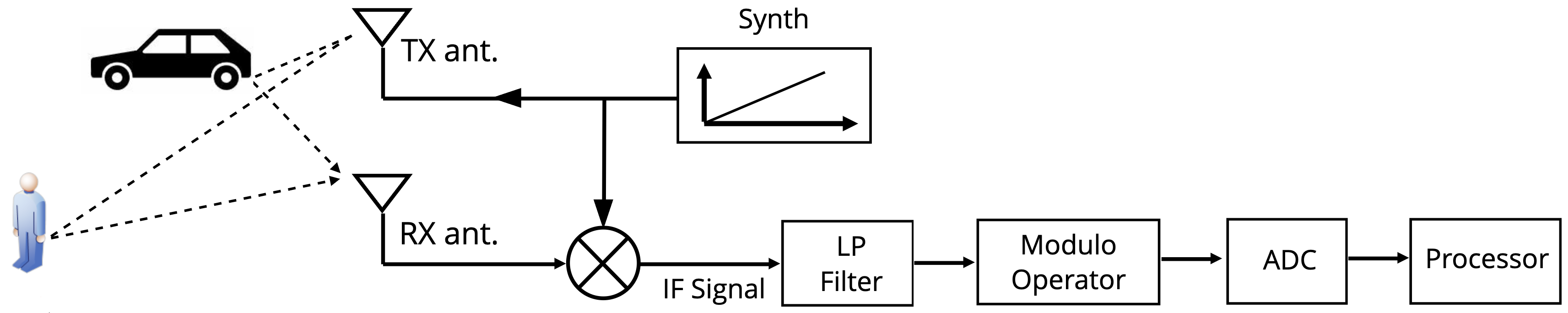}
		\caption{Range estimation using FMCW radar in the framework of unlimited sampling.}
		\label{FMCW_figure}
\end{figure*}

In this paper, we study the FMCW radar applied in range estimation problem where one transmit and one receive antenna pair is used in the context of unlimited sampling as shown in Fig. \ref{FMCW_figure}. This problem can be formulated as one-dimensional line spectral estimation (LSE) via unlimited sampling. In \cite{feuillen2022unlimited}, the Unlimited Sampling Framework (USF) is employed to address the problem, utilizing unlimited sampling ADCs at the hardware level and applying USAlg at the algorithmic level. However, it is worth noting that the original signal in \cite{feuillen2022unlimited} operates at a low noise level, indicating a relatively weak noise intensity compared to the signal. As previously mentioned, USAlg is particularly sensitive when noise is introduced to the original signal. In practice, the original signal is often affected by noise in many scenarios. Additionally, unlike situations in numerous papers where noise is added after modulo sampling, a noisy original signal not only influences the modulo values of samples but also affects the folding times. Therefore, studying robust reconstruction algorithms in the presence of noise before modulo sampling is of vital importance and is more challenging.

We study the LSE via US in the presence of noise before modulo sampling in this paper. The main contributions are threefold: Firstly, we establish an optimization problem to recover the folding instants and folding times. We show that the proposed objective function is meaningful even when the frequencies do not lie on the discrete Fourier transform (DFT) grid exactly, as the leakage onto a certain frequency band can be controlled. Secondly, an iterative algorithm named USLSE is proposed. Since the original optimization problem with quadratic objective function and integer constraints is NP-hard, we innovatively solve this problem by approximating the matrix in the quadratic form as a banded matrix. Such an approximation is reasonable as we show that the banded matrix consists of most of the energy of the original matrix, and the quadratic form with a banded matrix with integer constraints can be solved globally via dynamic programming (DP). In addition, provided that the estimates generated by the DP are close to the global solution, the residue calculated by subtracting the estimates from the global solution is sparse and orthogonal matching pursuit (OMP) is adopted to further refine the estimates. We could run multiple iterations of DP followed by OMP to refine the estimates. Finally, substantial numerical simulations and real experiments are conducted to demonstrate the robustness of the proposed USLSE compared to the state-of-the-art methods. Numerical results also demonstrate that the proposed approach is effective for recovering bandlimited signals.

\subsection{Notation}
We use $\mathbb{Z}$, $\mathbb{R}$, and $\mathbb{C}$ to denote sets of integers, real numbers, and complex numbers, respectively. Let $\mathbb{Z}^*$ be the set of complex numbers with both the real part and the imaginary part being integers. For a complex vector $\mathbf{x}\in\mathbb{C}^N$, $x_i$ or $[\mathbf{x}]_i$ denotes the $i$th element of $\mathbf{x}$ for any $i=1,\cdots,N$. For a complex matrix $\mathbf{A}\in\mathbb{C}^{N\times M}$, $A_{i,j}$ denotes the $(i,j)$th element of $\mathbf{A}$ for any $i=1,\cdots,N$ and $j = 1,\cdots, M$. Let ${\mathcal{S}}_N$ and ${\mathcal{S}}_M$ be subsets of $\{1,\cdots,N\}$ and $\{1,\cdots,M\}$, respectively. $|{\mathcal{S}}_N|$ denotes the cardinality of ${\mathcal{S}}_N$. $\mathbf{A}_{{\mathcal{S}}_N,{\mathcal{S}}_M}$ denotes the submatrix by choosing the rows indexed by $\mathcal{S}_N$ and the columns indexed by ${\mathcal{S}}_M$ of $\mathbf{A}$. For simplicity, $\mathbf{A}_{{\mathcal{S}}_N}$ is used to denote the submatrix by choosing the rows of $\mathbf{A}$ indexed by $\mathcal{S}_N$. Similarly, let $\mathbf{x}_{\mathcal{S}_N}$ denote the subvector by choosing the elements of $\mathbf{x}$ indexed by ${\mathcal{S}}_N$. Let $\Re\{\mathbf{A}\}$ and $\Im\{\mathbf{A}\}$ denote the real and imaginary parts of $\mathbf{A}$, respectively. Let $\operatorname{diag}(\mathbf{x})$ return a diagonal matrix with the diagonal being $\mathbf{x}$. Let $(\cdot)^*,(\cdot)^{\mathrm{T}}$ and $(\cdot)^{\mathrm{H}}$ be the conjugate, transpose, and Hermitian transpose operators, respectively. Let $\|\mathbf{x}\|_1, \|\mathbf{x}\|_2$ and $\|\mathbf{x}\|_{\infty}$ be the Manhattan norm, Euclidean norm, and maximum norm, respectively. Let $\odot$ denote the element-wise multiplication. Let $\lfloor\mathbf{A}\rfloor$ and $\lfloor\mathbf{A}\rceil$ be the element-wise floor and round operators for both real and imaginary parts of $\mathbf{A}$, respectively.

\section{Problem Setup}
In this section, the LSE via unlimited sampling is described. Consider the continuous LSE problem with $K$ frequencies
\begin{align}\label{conti_ori_obs}
g(t)=\sum\limits_{k=1}^Kc_k{\rm e}^{{\rm j}2\pi f_kt}+w(t),
\end{align}
where $f_k\in (0,f_{\rm max})$ and $c_k\in \mathbb{C}$ denote the frequency and amplitude of the $k$th signal, respectively. $t\in {\mathbb R}$ denotes the time, and $w(t)$ is the noise. Obviously, $f_{\rm max}$ is the Nyquist frequency, i.e., $f_{\rm Nyq}=f_{\rm max}$. Let $\gamma$ denote the oversampling factor with respect to the Nyquist frequency $f_{\rm Nyq}$. Thus, the sampling frequency is $f_{\rm s}=\gamma f_{\rm Nyq} = \gamma f_{\rm max}$ and the sampling interval is $T_{\rm s}=1/f_s = 1/(\gamma f_{\rm max})$. $g[n]\triangleq g(nT_{\rm s})$ are the original samples, which can be written as
\begin{align}\label{dis_ori_obs}
g[n]= x[n] + w[n], n = 0,\cdots, N-1,
\end{align}
where
\begin{align}\label{dis_ori_obs_xnoise}
    x[n] = \sum_{k=1}^Kc_k{\rm e}^{{\rm j}\omega_k n},
\end{align}
where $\omega_k \triangleq 2\pi f_k/f_s$ is the normalized angular frequency of the $k$th signal, $w[n]$ are the noise sampled from $w(t)$, and $N$ is the number of samples. In addition, for $f_k\in(0,f_{\rm max})$ and $f_s = \gamma f_{\rm max}$, we have
\begin{align}\label{omega_range}
    \omega_k = 2\pi f_k/f_s \in \left(0,\frac{2\pi}{\gamma}\right), k=1,\cdots,K.
\end{align}

Uniform modulo sampling of $g(t)$ where a modulo operator is applied before sampling yields
\begin{align}\label{fold_obs}
y[n]\triangleq \mathscr{M}_{\lambda}\left(\Re\left\{g[n]\right\}\right)+{\rm j}\mathscr{M}_{\lambda}\left(\Im\left\{g[n]\right\}\right),
\end{align}
where $\mathscr{M}_{\lambda}\left(t\right)$ is the centered modulo mapping defined as
\begin{align}
\mathscr{M}_{\lambda}\left(t\right)\triangleq2\lambda\left(\left\llbracket\frac{t}{2\lambda}+\frac{1}{2}\right\rrbracket-\frac{1}{2}\right),
\end{align}
$\llbracket t \rrbracket  \triangleq t-\lfloor t\rfloor$ denotes the fractional part of $t$, $\lambda$ denotes the threshold of ADC. Note that the original samples can be decomposed as the sum of modulo samples and a simple function \cite{bhandari2020unlimited}, i.e.,
\begin{align}\label{fold_obs_decom}
    g[n] = y[n] + 2\lambda\epsilon[n], n = 0,1,\cdots,N-1,
\end{align}
where $\epsilon[n]\in \mathbb{Z^*}$ and $\mathbb{Z}^* \triangleq \mathbb{Z} + {\rm j}\mathbb{Z}$ is the set of complex numbers with both real and imaginary parts being integers. Furthermore, the matrix representation of (\ref{dis_ori_obs}) and (\ref{fold_obs_decom}) can be formulated as
\begin{subequations}\label{prob_matrix}
\begin{align}
    &\mathbf{g} =\mathbf{x} + \mathbf{w} =\mathbf{A}(\boldsymbol{\omega})\mathbf{c} + \mathbf{w},\\
    &\mathbf{g} = \mathbf{y}+2\lambda\boldsymbol{\epsilon},
\end{align}
\end{subequations}
where $\mathbf{A}(\boldsymbol{\omega}) \triangleq [\mathbf{a}(\omega_1),\mathbf{a}(\omega_2),\cdots,\mathbf{a}(\omega_K)]\in\mathbb{C}^{N\times K}$ and  $\mathbf{a}(\omega_k) \triangleq \left[1, \mathrm{e}^{\mathrm{j} {\omega}_k}, \cdots, \mathrm{e}^{\mathrm{j}(N-1) {\omega}_k}\right]^{\mathrm{T}}$ is the array manifold steering vector. $\mathbf{g},\mathbf{x},\mathbf{w},\mathbf{y},\boldsymbol{\epsilon}$ are vector representations of the corresponding sequences. For example, $\mathbf{g} \triangleq [g[0],g[1],\cdots,g[N-1]]^{\rm T}$. $\boldsymbol{\omega} \triangleq [\omega_1,\cdots,\omega_K]^{\rm T}$ and $\mathbf{c} \triangleq [c_1,\cdots,c_K]^{\rm T}$. It is worth noting that $g_n = g[n-1], n = 1,\cdots,N$. In this paper, we aim to recover $\{\mathbf{c},\boldsymbol{\omega}\}$ from modulo samples ${\mathbf y}$.

To estimate parameters $\{\mathbf{c},\boldsymbol{\omega}\}$, least squares estimation is equivalent to solving the following optimization problem\footnote{If $\{w[n]\}_{n=0}^{N-1}$ are independent and identically distributed (i.i.d.) and follow a Gaussian distribution, then the least squares estimation is equivalent to maximum likelihood estimation.}
\begin{subequations}\label{MLE_opt}
\begin{align}
    &\underset{\mathbf{c},{\boldsymbol{\omega}},\boldsymbol{\epsilon}}{\operatorname{minimize}}~ \|\mathbf{y}+2\lambda\boldsymbol{\epsilon} - \mathbf{A}(\boldsymbol{\omega})\mathbf{c}\|_2^2
,\\
&{\rm s.t.}~\boldsymbol{\epsilon}\in {{\mathbb Z}^*}^N.
\end{align}
\end{subequations}
However, (\ref{MLE_opt}) is NP-hard generally as $\boldsymbol{\epsilon}$ are constrained on the lattice ${{\mathbb Z}^*}^N$. In detail, (\ref{MLE_opt}) can be solved in the following steps. Without loss of generality, suppose that $|\boldsymbol{\epsilon}|_{\infty}\triangleq \max\{\|\Re\{\boldsymbol\epsilon\}\|_{\infty},\|\Im\{\boldsymbol\epsilon\}\|_{\infty}\}=V_{\rm max}$ where $V_{\rm max}\in {\mathbb Z}$. We use enumeration to solve the problem. The number of possible $\boldsymbol\epsilon$ is $(2V_{\rm max}+1)^{2N}$. For each $\boldsymbol\epsilon$, (\ref{MLE_opt}) degenerates into a LSE which can be approximately solved and we can obtain the objective value. Finally, we choose the estimates of the parameters corresponding to the best (minimum) objective value. Consequently, the computation complexity is $(2V_{\rm max}+1)^{2N}$ times the complexity of the LSE algorithm. Note that the computation complexity is exponential in the number of measurements $N$, which makes it computationally intractable.

Another heuristic approach is to use a linear model to represent the nonlinear LSE model, i.e., the original $\mathbf{A}(\boldsymbol{\omega})\mathbf{c}$ can be represented as
\begin{align}\label{linear_repre}
\mathbf{A}(\boldsymbol{\omega})\mathbf{c} = \mathbf{A}_{\rm dic}\widetilde{\mathbf{c}},
\end{align}
where ${\mathbf A}_{\rm dic}\in {\mathbb C}^{N\times N_{\rm dic}}$ is the dictionary matrix with the $(n_1,n_2)$th element being ${\rm e}^{{\rm j}\frac{2\pi (n_1-1)(n_2-1)}{N_{\rm dic}}}$, $n_1=1,\cdots, N$, $n_2=1,\cdots, N_{\rm dic}$ and $N_{\rm dic}\geq N$. When $N_{\rm dic} \gg K$, $\widetilde{\mathbf{c}}$ is approximately sparse. Thus, (\ref{MLE_opt}) can be relaxed as
\begin{subequations}\label{MLE_opt_linear}
\begin{align}
    &\underset{\tilde{\mathbf{c}},\boldsymbol{\epsilon}}{\operatorname{minimize}}~ \|\mathbf{y}+2\lambda\boldsymbol{\epsilon} - \mathbf{A}_{\rm dic}\widetilde{\mathbf{c}}\|_2^2+\mu\|\widetilde{\mathbf{c}}\|_1
,\\
&{\rm s.t.}~\boldsymbol{\epsilon}\in {{\mathbb Z}^*}^N,\label{constraintlattice}
\end{align}
\end{subequations}
where $\mu$ is a regularization parameter to control the sparsity of the signal. Problem (\ref{MLE_opt_linear}) is still NP-hard as there exist integer constraints that $\boldsymbol{\epsilon}\in {{\mathbb Z}^*}^N$. We solved the relaxed version of (\ref{MLE_opt_linear}) by relaxing the non-convex constraints (\ref{constraintlattice}) as the linear inequality constraints, but the solution is poor. In particular, if we ignore the integer constraints first, the solution of (\ref{MLE_opt_linear}) is trivial and far from the true values, i.e., $\widehat{\boldsymbol{\epsilon}} = -\mathbf{y}/(2\lambda)$ and $\widehat{\widetilde{\mathbf{c}}} = \mathbf{0}$. Besides, we kept the non-convex constraints (\ref{constraintlattice}) and used the proximal approach to solve (\ref{MLE_opt_linear}) with a random initial point. However, the local solution is still poor. Therefore, it is challenging to solve (\ref{MLE_opt_linear}) using polynomial time algorithms, and more structures should be utilized.

\section{LSE through Frequency Domain Processing}
Intuitively, oversampling benefits in two facets: First, oversampling makes the simple function $2\lambda\boldsymbol{\epsilon}$ similar at the nearby instants. This causes the value of the first-order finite difference of the simple function to be near zero, which reduces the state space of the first-order finite difference. Second, oversampling widens the spectrum between the maximum signal frequency and the Nyquist frequency. Since the leakage due to the line spectral signal into the spectrum between the maximum signal frequency and the sampling frequency is small which will be discussed below, the spectrum between the maximum signal frequency and the sampling frequency is only approximately related to the simple function. This leads us to expect that the recovery process becomes simpler. In the following, we exploit the above benefits to formulate the LSE.

Let $\underline{\boldsymbol{\epsilon}}$ denote the first-order finite difference of ${\boldsymbol{\epsilon}}$, where the $n$th element is defined as
\begin{align}
\underline{{\epsilon}}_n=
    {\epsilon}_{n+1}-{\epsilon}_n,~n=1,\cdots,N-1,
\end{align}
or in matrix form $\underline{{\boldsymbol{\epsilon}}}={\mathbf J}\boldsymbol{\epsilon}$ where ${\mathbf J} = \mathbf{I}_{2:N,:} - \mathbf{I}_{1:N-1,:}$ and $\mathbf{I}\in\mathbb{R}^{N\times N}$ is the identity matrix. Note that $\underline{\boldsymbol{\epsilon}}$ is still on the lattice ${{\mathbb Z}^*}^{N-1}$, thus $\Re\{\underline{\epsilon}_n\}$ and $\Im\{\underline{\epsilon}_n\}$ will be in the set $\{-V,\cdots,0,\cdots, V\}$ where $V\triangleq\max\{\|\Re\{\boldsymbol{\underline{\epsilon}}\}\|_{\infty},\|\Im\{\boldsymbol{\underline{\epsilon}}\}\|_{\infty}\}$ is a positive integer. As the oversampling factor $\gamma$ increases, the dynamic range of both the real and imaginary parts of $\underline{\boldsymbol{\epsilon}}$, i.e.,$V$, will decrease, and Proposition \ref{propdepsbound} shows the result of the noiseless case. Applying the first-order finite difference to (\ref{prob_matrix}), we have
\begin{align}\label{diff_decom}
\underline{\mathbf y}+2\lambda\underline{\boldsymbol\epsilon} = \underline{\mathbf{x}} + \underline{\mathbf{w}}=\mathbf{J} \mathbf{A}(\boldsymbol{\omega}){\mathbf{c}} +  \underline{\mathbf{w}},
\end{align}
where $\underline{\mathbf{y}} \triangleq \mathbf{J}\mathbf{y}$, $\underline{\mathbf{x}} \triangleq \mathbf{J}\mathbf{x}$, and $\underline{\mathbf{w}} \triangleq \mathbf{J}\mathbf{w}$.
\begin{proposition}\label{propdepsbound}
In the noiseless case, given the oversampling factor $\gamma$, both $\Re\{\underline{\epsilon}_n\}$ and $\Im\{\underline{\epsilon}_n\}$, $n=1,2,\cdots,N-1$, will be in the set $\{-V,\cdots,0,\cdots, V\}$ with a bounded value of $V$ such that
\begin{align}
    V\leq \left\lfloor\frac{K\pi c_{\rm max}}{\lambda\gamma} + 1\right\rfloor,
\end{align}
where $c_{\rm max}=\underset{k}{\operatorname{max}}|c_k|$.
\end{proposition}
\begin{proof}
The proof is postponed to Appendix \ref{state_space_est}.
\end{proof}

Performing DFT on (\ref{diff_decom}) yields
\begin{align}\label{fft_diff_decom}
   \widetilde{\underline{\mathbf y}}+2\lambda\mathbf{F}\underline{\boldsymbol\epsilon}=\widetilde{\underline{\mathbf x}} + \widetilde{\underline{\mathbf{w}}} = \mathbf{F}\mathbf{J}\mathbf{A}(\boldsymbol{\omega})\mathbf{c}+ \widetilde{\underline{\mathbf{w}}},
\end{align}
where $\widetilde{\underline{\mathbf y}}\triangleq{\mathbf F}\underline{\mathbf y}$, $\widetilde{\underline{\mathbf x}}\triangleq{\mathbf F}\underline{\mathbf x}$, $\widetilde{\underline{\mathbf{w}}}\triangleq {\mathbf F}\underline{\mathbf w}$, and $\mathbf{F}\in\mathbb{C}^{(N-1)\times(N-1)}$ is the DFT matrix with the $(n_1,n_2)$th element being $\frac{1}{\sqrt{N-1}}{\rm e}^{-{\rm j}\frac{2\pi (n_1-1)(n_2-1)}{N-1}}$. Given that $\omega_k \leq 2\pi/\gamma$ for $k=1,\cdots, K$ as stated in (\ref{omega_range}), if $\omega_k$ lies on the grid $\{\frac{2\pi n}{N-1}\}_{n=0}^{N-2}$ for all $k$, the $(n+1)$th element of $\widetilde{\underline{\mathbf{x}}}$ is
\begin{align}
   \widetilde{\underline{{x}}}[n] &\triangleq \widetilde{\underline{\mathbf{x}}}_{n+1} = \mathbf{F}_{n+1}\mathbf{J}\mathbf{A}(\boldsymbol{\omega})\mathbf{c}\notag\\&= \mathbf{F}_{n+1}[\mathbf{A}(\boldsymbol{\omega})]_{1:N-1,:}(\mathbf{D} - \mathbf{I}_K)\mathbf{c}\notag\\
   &=[0,0,\cdots,0](\mathbf{D} - \mathbf{I}_K)\mathbf{c}= 0
\end{align}
for $n> \lfloor\frac{N-1}{\gamma}\rfloor$, where $\mathbf{D} = {\rm diag}({\rm e}^{{\rm j}\omega_1},\cdots,{\rm e}^{{\rm j}\omega_K})$. When frequencies $\boldsymbol{\omega}$ are not on the grid, despite the presence of spectral leakage, it can be expected that the magnitude of the $(n+1)$th element of the first-order difference line spectral signal in the frequency domain, i.e., $|\widetilde{\underline{{x}}}_{n+1}|$ or $|\widetilde{\underline{{x}}}[n]|$, is small for some $n > \lfloor\frac{N-1}{\gamma}\rfloor$. In the following, a formal analysis is provided. We first derive the upper bound of the magnitude of the $(n+1)$th element of $\widetilde{\underline{\mathbf x}}$ which is shown in Proposition \ref{propupperx}. For $(N-1)(\frac{1}{\gamma}+\beta) \leq n\leq (N-1)(1-\beta)$ where $\beta$ is a constant factor such that $\frac{1}{N-1}<\beta < \frac{\gamma-1}{2\gamma}$, according to (\ref{upperx}), we have 
\begin{align}\label{proposition2extend}
    |\underline{\widetilde{{x}}}[n]| &\leq \frac{2K\pi c_{\rm max}s_{\rm max}\delta_{\rm max}}{\sqrt{N-1}\min\{\sin(\frac{\pi n}{N-1}-\frac{\pi}{\gamma}), \sin(\frac{\pi n}{N-1})\}}\notag\\
    &\stackrel{a}\leq \frac{2K\pi c_{\rm max}s_{\rm max}\delta_{\rm max}}{\sqrt{N-1}\min\{\sin(\pi(\frac{1}{\gamma} + \beta)), \sin(\beta\pi)\}}\notag\\
    &\leq \frac{2K\pi c_{\rm max}s_{\rm max}\delta_{\rm max}}{\sqrt{N-1}\sin(\beta\pi)},
\end{align}
where $\stackrel{a}\leq$ is due to both $\sin(\frac{\pi n}{N-1}-\frac{\pi}{\gamma})$ and $\sin(\frac{\pi n}{N-1})$ are greater than $0$ and are concave for $(N-1)(\frac{1}{\gamma}+\beta) \leq n\leq (N-1)(1-\beta)$. Thus, $|\underline{\widetilde{{x}}}[n]|$ can be arbitrarily small with increasing $N$. This demonstrates that with enough samples, one could choose the samples $\widetilde{\underline{y}}[n]$ with index $\{\lfloor (N-1)(\frac{1}{\gamma}+\beta)\rfloor+1,\lfloor (N-1)(\frac{1}{\gamma}+\beta)\rfloor+2,\cdots,\lfloor (N-1)(1-\beta)\rfloor\}$ such that the leakage is negligible, thus an optimization problem only with respect to $\underline{\boldsymbol{\epsilon}}$ can be established which is introduced below.

\begin{proposition}\label{propupperx}
Define $a_k\in\mathbb{Z}$, $a_k\leq \lfloor\frac{N-1}{\gamma}\rfloor$ and $\delta_k\in[-0.5,0)\cup(0,0.5)$, $k=1,\cdots,K$ such that $\omega_k = \frac{2\pi}{N-1}(a_k+\delta_k)$. For $n>\lfloor\frac{N-1}{\gamma}\rfloor$, the magnitude of the $(n+1)$th element of $\widetilde{\underline{\mathbf x}}$ can be upper bounded by
\begin{align}\label{upperx}
    |\underline{\widetilde{{x}}}[n]| \leq \frac{2K\pi c_{\rm max}s_{\rm max}\delta_{\rm max}}{\sqrt{N-1}\min\{\sin(\frac{\pi n}{N-1}-\frac{\pi}{\gamma}), \sin(\frac{\pi n}{N-1})\}},
\end{align}
where $c_{\rm max} \triangleq \operatorname{max}_{i=1}^K|c_i|$, $s_{\rm max}\triangleq \max_{i=1}^K\sin(\frac{\omega_k}{2})$, and $\delta_{\rm max} \triangleq \operatorname{max}_{i=1}^K|\delta_i|$.
\end{proposition}
\begin{proof}
The proof is postponed to Appendix \ref{propupperxapp}.
\end{proof}

Let $\mathcal{S}$ be a subset of $\{1,2,\cdots,N-1\}$ defined as 
\begin{align}\label{subset_M}
    \mathcal{S} \triangleq \{\lfloor \frac{N-1}{\gamma}+N_{\beta}\rfloor+2,\cdots,\lfloor N-1-N_{\beta}\rfloor+1\},
\end{align}
where $N_{\beta}\triangleq (N-1)\beta$ and $\frac{1}{N-1}<\beta<\frac{\gamma-1}{2\gamma}$ is a constant factor. By choosing the $\mathcal{S}$ elements of (\ref{fft_diff_decom}), a submodel
\begin{align}\label{F_delta_g_sub}
    \widetilde{\underline{\mathbf y}}_{\mathcal{S}}+2\lambda\mathbf{F}_{\mathcal{S}}\underline{\boldsymbol\epsilon}= \widetilde{\underline{\mathbf{x}}}_{\mathcal{S}} + \widetilde{\underline{\mathbf{w}}}_{\mathcal{S}}
\end{align}
can be obtained. Note that the number of measurements of model (\ref{F_delta_g_sub}) is $|\mathcal{S}| = \lfloor N-1-N_{\beta}\rfloor - \lfloor \frac{N-1}{\gamma}+N_{\beta}\rfloor\approx (N-1)(1-\frac{1}{\gamma}-2\beta)$. On the one hand, a smaller $\beta$ means that more data can be used to perform signal recovery, and the measurement matrix $2\lambda\mathbf{F}_{\mathcal{S}}$ has a lower mutual coherence. On the other hand, as $\beta$ increases, the impact caused by spectrum leakage will be reduced, as stated in Proposition \ref{propupperx} and (\ref{proposition2extend}). Thus, there exists a tradeoff for selecting $\beta$. We validated this phenomenon through simulation for the algorithm we proposed later on in Section \ref{simu_M}. By treating the right-hand side of (\ref{F_delta_g_sub}), i.e., $\widetilde{\underline{\mathbf{x}}}_{\mathcal{S}} + \widetilde{\underline{\mathbf{w}}}_{\mathcal{S}}$, as noise, based on the least squares method, the subproblem of estimating $\underline{\boldsymbol{\epsilon}}$ can be formulated as
\begin{subequations}\label{opt_diff_sub}
\begin{align}
&\underset{\underline{\boldsymbol{\epsilon}}}{\operatorname{minimize}}~ \|{\mathbf{z}}_{\mathcal{S}}+\mathbf{F}_{\mathcal{S}}\underline{\boldsymbol{\epsilon}}\|_2^2
,\\
&{\rm s.t.}~\underline{\epsilon}_n\in\mathcal{V}, n =1,\cdots,N-1, 
\end{align}	
\end{subequations}
where $\mathbf{z}_{\mathcal{S}} \triangleq\widetilde{\underline{\mathbf{y}}}_{\mathcal{S}}/(2\lambda)$, $\mathcal{V} = \{m+{\rm j}n|-V\leq m,n\leq V,m,n\in\mathbb{Z}\}$ is the bounded lattice and $V$ is a small integer as shown in Proposition \ref{propdepsbound}. Problem (\ref{opt_diff_sub}) is an integer quadratic programming problem which is also NP-hard generally. In the following, an efficient algorithm based on DP is proposed to approximately solve (\ref{opt_diff_sub}). 
\section{Algorithm}
This section provides a polynomial-time algorithm for solving (\ref{opt_diff_sub}) approximately. In general, this problem is difficult to solve. Our approach consists of two steps: Firstly, the original problem (\ref{opt_diff_sub}) is solved approximately by exploiting the fine structure of (\ref{opt_diff_sub}). That is, we approximate the original problem (\ref{opt_diff_sub}) in a reasonable way and use the DP method to solve the problem globally. Secondly, the approximate solution acts as a good candidate, and heuristic approaches are used to further refine the quality of the solution by solving the original problem (\ref{opt_diff_sub}). At the end of this section, we will provide the validation of the proposed algorithm by showing the rationality of solving the approximate optimization problem via DP.

\subsection{Dynamic Programming Method}
By expanding the quadratic term, problem (\ref{opt_diff_sub}) can be rewritten as
\begin{subequations}\label{opt_diff_sub_re}
\begin{align}
    &\underset{\underline{\boldsymbol{\epsilon}}}{\operatorname{minimize}}~ \underline{\boldsymbol{\epsilon}}^{\rm H}\mathbf{Q}\underline{\boldsymbol{\epsilon}} + 2\Re\{\mathbf{b}^{\rm H}\underline{\boldsymbol{\epsilon}}\}
,\\
&{\rm s.t.}~\underline{\epsilon}_n\in\mathcal{V}, n =1,\cdots,N-1,
\end{align}	
\end{subequations}
where $\mathbf{Q} = \mathbf{F}_{\mathcal{S}}^{\rm H}\mathbf{F}_{\mathcal{S}}$ and $\mathbf{b} = \mathbf{F}_{\mathcal{S}}^{\rm H}{{\mathbf{z}}}_{\mathcal{S}}$. Note that the $(i,j)$th term of $\mathbf{Q}$ is
\begin{align}\label{Qij}
    Q_{i,j} & = \mathbf{F}^{\rm H}_{\mathcal{S},i}\mathbf{F}_{\mathcal{S},j}.
\end{align}
When $\mathcal{S} = \{1,\cdots, N-1\}$, $\mathbf{F}_{\mathcal{S}} = \mathbf{F}$ is the DFT matrix. Thus, $\mathbf{F}_{\mathcal{S},i}$ and $\mathbf{F}_{\mathcal{S},j}$ are orthogonal when $i\neq j$ and $\mathbf{F}^{\rm H}_{\mathcal{S},i}\mathbf{F}_{\mathcal{S},j} = 1$ when $i=j$ and $\mathbf{Q}$ is an identity matrix. For $\mathcal{S}$ defined in (\ref{subset_M}) which is a subset of $\{1,\cdots, N-1\}$, although $\mathbf{Q}$ is not an identity matrix, elements with relatively large magnitudes still concentrate near the diagonal. In detail, when $i=j$, the magnitude of $Q_{i,j}$ will get closer to $1$ as $|\mathcal{S}|$ increases. When $i\neq j$, the magnitude of $Q_{i,j}$ will be relatively small, especially when the wrap-around distance between $i$ and $j$, i.e., $\min\{|i-j|,N-1-|i-j|\}$, is far away from $0$. Thus, the matrix $\mathbf{Q}$ only has relatively large values on its diagonal and a few adjacent positions (wrap-around distance). Approximating $\mathbf{Q}$ as a $p$th order diagonal matrix $\widetilde{\mathbf{Q}}$ is reasonable, where $\widetilde{\mathbf{Q}}$ is a submatrix of $\mathbf{Q}$ and is defined as follows
\begin{align}\label{Q_tilde}
        \widetilde{Q}_{i,j} = \begin{cases}
        Q_{i,j}~ |i-j|\leq p, \\
        0,~\text{otherwise},
    \end{cases}
\end{align}
where $i = 1,\cdots,N-1$ and $j = 1,\cdots,N-1$. Note that $\widetilde{\mathbf{Q}}$ is still a Hermitian matrix. In Section \ref{DP_approx_well}, a detailed analysis of the rationality of approximating ${\mathbf{Q}}$ as ${\widetilde{\mathbf{Q}}}$ will be provided. By replacing $\mathbf{Q}$ with $\widetilde{\mathbf{Q}}$ in (\ref{opt_diff_sub_re}), an approximate optimization problem can be formulated as
\begin{subequations}\label{opt_diff_sub_appro}
\begin{align}
    &\underset{\underline{\boldsymbol{\epsilon}}}{\operatorname{minimize}}~ \underline{\boldsymbol{\epsilon}}^{\rm H}\widetilde{\mathbf{Q}}\underline{\boldsymbol{\epsilon}} + 2\Re\{\mathbf{b}^{\rm H}\underline{\boldsymbol{\epsilon}}\}
,\\
&{\rm s.t.}~\underline{\epsilon}_n\in\mathcal{V}, n =1,\cdots,N-1.
\end{align}	
\end{subequations}
We first consider the quadratic term in (\ref{opt_diff_sub_appro}), i.e., $\underline{\boldsymbol{\epsilon}}^{\rm H}\widetilde{\mathbf{Q}}\underline{\boldsymbol{\epsilon}}$. For $\widetilde{Q}_{i,j} = 0$ when $|i-j| > p$, the cost function  $\underline{\boldsymbol{\epsilon}}^{\rm H}\widetilde{\mathbf{Q}}\underline{\boldsymbol{\epsilon}}$ is a function of
$\underline{\epsilon}_n$  and its nearby terms $\{\underline{\epsilon}_q,0<|q-n|\leq p \}$. In addition, the second term in (\ref{opt_diff_sub_re}) can be decomposed as independent sums, i.e., $2\Re\{\mathbf{b}^{\rm H}\underline{\boldsymbol{\epsilon}}\} = \sum_{i=1}^{N-1}2\Re\{b_i^{*}\}\underline{{\epsilon}}_i$. Thus, parameters $\underline{\boldsymbol{\epsilon}}$ in problem (\ref{opt_diff_sub_appro}) have a $p$th order Markov property, and the global optimum can be obtained by using a DP method.

Let $h(\underline{\boldsymbol{\epsilon}}) \triangleq \underline{\boldsymbol{\epsilon}}^{\rm H}\widetilde{\mathbf{Q}}\underline{\boldsymbol{\epsilon}} + 2\Re\{\mathbf{b}^{\rm H}\underline{\boldsymbol{\epsilon}}\}$ which is the objective function of problem (\ref{opt_diff_sub_appro}), and we decompose it into $P \triangleq N-p-1$ terms, i.e.,
\begin{align}\label{decom_appr_loss_1}
     h(\underline{\boldsymbol{\epsilon}}) = \underbrace{\sum_{n=1}^{P-1} \underline{\boldsymbol{\epsilon}}^{\rm H}\widetilde{\mathbf{Q}}^{(n)}\underline{\boldsymbol{\epsilon}} + 2\Re\{{b}_n^*\underline{{\epsilon}}_n\}}_{\text{first $P-1$ terms}} + \underbrace{\Bigg(\underline{\boldsymbol{\epsilon}}^{\rm H}\widetilde{\mathbf{Q}}^{(P)}\underline{\boldsymbol{\epsilon}} + 2\Re\{\mathbf{b}_{P:N-1}^{\rm H}\underline{{\boldsymbol{\epsilon}}}_{P:N-1}\}\Bigg)}_{\text{the $P$th term}}
\end{align}
where $\widetilde{\mathbf{Q}} = \sum_{n=1}^{P}\widetilde{\mathbf{Q}}^{(n)}$ and $\widetilde{\mathbf{Q}}^{(n)}$ are all Hermitian matrices defined as follows. For $n = 1,\cdots, P-1$, $\widetilde{\mathbf{Q}}^{(n)}\in \mathbb{C}^{(N-1)\times (N-1)}$ is a submatrix of $\widetilde{\mathbf{Q}}$ retaining the elements in columns $n$ through $N-1$ of $\widetilde{\mathbf{Q}}$'s $n$th row, as well as the elements in rows $n$ through $N-1$ of $\widetilde{\mathbf{Q}}$'s $n$th column defined as
\begin{align}
    \widetilde{Q}^{(n)}_{i,j}=\begin{cases}
        \widetilde{Q}_{i,j}~ (i=n \text{ and } j\geq i) \text{ or } (j=n \text{ and } i\geq j), \\
        0,~\text{otherwise}.
    \end{cases}\notag
\end{align}
For the last term, $\widetilde{\mathbf{Q}}^{(P)}\in  \mathbb{C}^{(N-1)\times (N-1)}$ is a submatrix of $\widetilde{\mathbf{Q}}$ retaining both $\{P,\cdots,N-1\}$ rows and columns of $\widetilde{\mathbf{Q}}$ defined as
\begin{align}\label{Q_last_P}
    \widetilde{Q}^{(P)}_{i,j}=\begin{cases}
    \widetilde{Q}_{i,j}~ (i\geq P) \text{ or } (j\geq P), \\
    0,~\text{otherwise}.
    \end{cases}
\end{align}
Because $\widetilde{Q}_{i,j}=0$ when $|i-j|>p$ (\ref{Q_tilde}), we have $\widetilde{Q}_{n,j}^{(n)}= 0$ for $j>n+p$ and $\widetilde{Q}_{i,n}^{(n)}= 0$ for $i>n+p$. Thus, the $n$th term, $n=1,\cdots, P-1$, of the decomposed loss (\ref{decom_appr_loss_1}) can be simplified as
\begin{align}\label{h_n}
    {\underline{\boldsymbol{\epsilon}}^{\rm H}\widetilde{\mathbf{Q}}^{(n)}\underline{\boldsymbol{\epsilon}} + 2\Re\{{b}_n^*\underline{{\epsilon}}_n\}}
    = \underline{\epsilon}_{n}^*\widetilde{Q}_{n,n}\underline{\epsilon}_{n} + 2\Re\{\underline{\epsilon}_{n}^*\widetilde{\mathbf{Q}}_{n,n+1:n+p}\underline{\boldsymbol{\epsilon}}_{n+1:n+p}\} + 2\Re\{{b}_n^*\underline{{\epsilon}}_n\}\triangleq  h_{n}(\underline{\boldsymbol{\epsilon}}_{n:n+p})
\end{align}
which is only related to variables $\underline{\boldsymbol{\epsilon}}_{n:n+p}$ and denoted as $ h_{n}(\underline{\boldsymbol{\epsilon}}_{n:n+p})$. For the last term in (\ref{decom_appr_loss_1}), according to (\ref{Q_last_P}), we have
\begin{align}\label{h_n_last}
    \underline{\boldsymbol{\epsilon}}^{\rm H}\widetilde{\mathbf{Q}}^{(P)}\underline{\boldsymbol{\epsilon}} + 2\Re\{\mathbf{b}_{P:N-1}^{\rm H}\underline{{\boldsymbol{\epsilon}}}_{P:N-1}\}
    = \underline{\boldsymbol{\epsilon}}_{P:N-1}^{\rm H}\widetilde{\mathbf{Q}}_{P:N-1,P:N-1} \underline{\boldsymbol{\epsilon}}_{P:N-1} + 2\Re\{\mathbf{b}_{P:N-1}^{\rm H}\underline{{\boldsymbol{\epsilon}}}_{P:N-1}\}
    \triangleq h_{P}(\underline{\boldsymbol{\epsilon}}_{P:N-1}),
\end{align}
which is only related to the last $p+1$ variables $\underline{\boldsymbol{\epsilon}}_{P:N-1}$ and denoted as $ h_{P}(\underline{\boldsymbol{\epsilon}}_{P:N-1})$. In summary, we can decompose the objective function (\ref{decom_appr_loss_1}) into $P$ terms, i.e.,
\begin{align}\label{decom_appr_loss_2}
    h(\underline{\boldsymbol{\epsilon}}) = \sum_{n=1}^{P}h_{n}(\underline{\boldsymbol{\epsilon}}_{n:n+p}).
\end{align}
where $h_{n}(\underline{\boldsymbol{\epsilon}}_{n:n+p})$ are defined in (\ref{h_n}) and (\ref{h_n_last}).

Below the DP is utilized to solve the approximate optimization problem (\ref{opt_diff_sub_appro}) by using the decomposed form of the objective function $h(\underline{\boldsymbol{\epsilon}})$ (\ref{decom_appr_loss_2}). We first define the optimization problem at stage $n$ (for the first $n$ variables) as
\begin{align}\label{dp_prob}
    {\rm DP}_n(\underline{{\boldsymbol{\epsilon}}}_{n+1:n+p}) = \underset{\underline{\boldsymbol{\epsilon}}_{1:n}\in\mathcal{V}^{n}}{\operatorname{minimize}}~ \sum_{i=1}^{n} h_i(\underline{{\boldsymbol{\epsilon}}}_{i:i+p}),
\end{align}
$n = 1,\cdots,P-1$. Observe that the optimization problem (\ref{opt_diff_sub_appro}) is equivalent to
\begin{subequations}\label{last_DP_opt}
\begin{align}
    &\underset{\underline{\boldsymbol\epsilon}_{P:N-1}}{\operatorname{minimize}}~ {\rm DP}_{P-1}(\underline{{\boldsymbol{\epsilon}}}_{P:N-2}) + h_{P}(\underline{{\boldsymbol{\epsilon}}}_{P:N-1}),\\
&{\rm s.t.}~\underline{\epsilon}_n\in\mathcal{V}, n =1,\cdots,N-1,
\end{align}
\end{subequations}
which can be viewed as the optimization problem at the last stage (stage $P$). It also should be noted that we can recursively solve $ {\rm DP}_n(\underline{{\boldsymbol{\epsilon}}}_{n+1:n+p})$ as follows
\begin{align}\label{DP_recur}
    {\rm DP}_n(\underline{{\boldsymbol{\epsilon}}}_{n+1:n+p}) 
    = \underset{\underline{\epsilon}_{n}\in\mathcal{V}}{\operatorname{minimize}}~ {\rm DP}_{n-1}(\underline{{\boldsymbol{\epsilon}}}_{n:n+p-1}) + h_n(\underline{{\boldsymbol{\epsilon}}}_{n:n+p}),
\end{align}
$n = 1,\cdots,P-1$, where $ {\rm DP}_0(\underline{{\boldsymbol{\epsilon}}}_{1:p}) = 0$ is the initialization term. For $\underline{\epsilon}_n\in\mathcal{V}, n=1,\cdots, N-1$, the enumeration method is utilized here to solve the recursion (\ref{DP_recur}), and the computational complexity is $O(p^2|\mathcal{V}|^{p+1})$. In addition, the optimal solution $\underline{\widehat{\epsilon}}_{n}$ will be recorded by ${\rm RC}_{n}(\underline{{\boldsymbol{\epsilon}}}_{n+1:n+p})$ for $n =1,\cdots, P-1$. Finally, we will solve the final optimization problem (\ref{last_DP_opt}) based on ${\rm DP}_{P-1}(\underline{{\boldsymbol{\epsilon}}}_{P:N-2})$. Similarly, an enumeration method can be applied to obtain the optimal solution $\widehat{\underline{\boldsymbol\epsilon}}_{P:N-1}$, and the computational complexity of the final step is $O(p^2|\mathcal{V}|^{p+1})$. To obtain the corresponding optimal variables, a backward process
\begin{align}
    \widehat{\underline{\epsilon}}_{n} = {\rm RC}_{n}(\widehat{\underline{\boldsymbol{\epsilon}}}_{n+1:n+p})
\end{align}
is recursively executed, for $n = P-1,\cdots, 1$. The initial estimates $\widehat{\underline{\boldsymbol{\epsilon}}}_{P:N-1}$ is the solution of (\ref{last_DP_opt}). The DP algorithm is summarized in Algorithm \ref{DP_alg}.
\begin{algorithm}[ht]
\caption{Dynamic Programming Algorithm}\label{DP_alg}
\begin{algorithmic}[1]
\STATE \textbf{Inputs}: Observations $\mathbf{z}_{\mathcal{S}}$, measurement matrix $\mathbf{F}_{\mathcal{S}}$, state space $\mathcal{V}$, and order $p$.
\STATE Compute $\mathbf{b} = \mathbf{F}_{\mathcal{S}}^{\rm H}\mathbf{z}_{\mathcal{S}}$ and $\widetilde{\mathbf{Q}}^{(n)}, n= 1,\cdots, P$.
\STATE \textbf{Forward process:}
\STATE  Initialize ${\rm DP}_0(\underline{{\boldsymbol{\epsilon}}}_{1:p}) = 0$.
\FOR{$n=1,\cdots,P-1$}
\STATE Compute $ {\rm DP}_n(\underline{{\boldsymbol{\epsilon}}}_{n+1:n+p})$ according to (\ref{DP_recur}) and store it, and delete $ {\rm DP}_{n-1}(\underline{{\boldsymbol{\epsilon}}}_{n:n+p-1})$.
\STATE Store the corresponding optimal solution $\widehat{\underline{\epsilon}}_n$ in ${\rm RC}_{n}(\underline{{\boldsymbol{\epsilon}}}_{n+1:n+p})$.
\ENDFOR
\STATE Obtain the estimates $\widehat{\underline{\boldsymbol\epsilon}}_{P:N-1}$ by solving (\ref{last_DP_opt}), and  delete $ {\rm DP}_{P-1}(\underline{{\boldsymbol{\epsilon}}}_{P:N-2})$.
\STATE \textbf{Backward process:}
\FOR{$n=P-1,\cdots,1$}
\STATE Compute $\widehat{\underline{\epsilon}}_{n} = {\rm RC}_{n}(\widehat{\underline{\boldsymbol{\epsilon}}}_{n+1:n+p})$, and delete ${\rm RC}_{n}(\widehat{\underline{\boldsymbol{\epsilon}}}_{n+1:n+p})$.
\ENDFOR
\STATE \textbf{Outputs:} $\widehat{\underline{\boldsymbol{\epsilon}}}$.
\end{algorithmic}
\end{algorithm}
\subsection{USLSE Algorithm}
The DP method has been used to solve the problem (\ref{opt_diff_sub_re}) approximately. Here we propose an enhanced iterative algorithm to further refine the solution $\widehat{\underline{\boldsymbol{\epsilon}}}$. The algorithm proceeds as follows:
By viewing $\widehat{\underline{\boldsymbol{\epsilon}}}$ as the initial value, $\underline{\boldsymbol{\epsilon}}$ can be decomposed into two parts, i.e., $\underline{\boldsymbol{\epsilon}} = \widehat{\underline{\boldsymbol{\epsilon}}} + \Delta\underline{\boldsymbol{\epsilon}}$, and the original optimization problem (\ref{opt_diff_sub}) is
\begin{subequations}\label{opt_diff_sub_sparse}
\begin{align}
    &\underset{\Delta\underline{\boldsymbol{\epsilon}}}{\operatorname{minimize}}~ \|\mathbf{z}_{\mathcal{S}}+\mathbf{F}_{\mathcal{S}}\widehat{\underline{\boldsymbol{\epsilon}}} + \mathbf{F}_{\mathcal{S}}\Delta\underline{\boldsymbol{\epsilon}}\|_2^2
,\\
&{\rm s.t.}~\Delta\underline{\epsilon}_{n}\in \mathbb{Z}^*, n =1,\cdots,N-1.
\end{align}	
\end{subequations}
Given that $\widehat{\underline{\boldsymbol{\epsilon}}}$ acts as a good approximation to the optimal solution, $\Delta\underline{\boldsymbol{\epsilon}}$ will be a sparse vector. Thus, problem (\ref{opt_diff_sub_sparse}) can be viewed as a sparse recovery problem, and methods like OMP or basis pursuit can be applied. Here, for $\Delta\underline{\epsilon}_{n}\in \mathbb{Z}^*$, we will run OMP algorithm, and for each iteration, the estimate will be rounded to the lattice $\mathbb{Z}^*$. Finally, the estimates will be updated as $\widehat{\underline{\boldsymbol{\epsilon}}}:= \widehat{\underline{\boldsymbol{\epsilon}}} + \Delta\widehat{\underline{\boldsymbol{\epsilon}}}$ provided that the refined estimate decreases the objective function value, otherwise we still use the old estimate. We term the above procedure as ``DP+OMP''. 

Multiple iterations of ``DP+OMP'' can be executed to enhance performance. Given the estimate $\widehat{\underline{\boldsymbol{\epsilon}}}$, we expand (\ref{opt_diff_sub_sparse}) and obtain
\begin{subequations}\label{opt_diff_sub_sparse_expand}
\begin{align}
    &\underset{\Delta\underline{\boldsymbol{\epsilon}}}{\operatorname{minimize}}~(\Delta\underline{\boldsymbol{\epsilon}})^{\rm H}{\mathbf Q}\Delta\underline{\boldsymbol{\epsilon}}+2\Re\{(\mathbf{z}_{\mathcal{S}}+\mathbf{F}_{\mathcal{S}}\widehat{\underline{\boldsymbol{\epsilon}}})^{\rm H}\mathbf{F}_{\mathcal{S}}\Delta{\underline{\boldsymbol{\epsilon}}}\}
,\\
&{\rm s.t.}~\Delta\underline{\epsilon}_{n}\in \mathbb{Z}^*, n =1,\cdots,N-1.
\end{align}	
\end{subequations}
We still use the same banded matrix $\widetilde{\mathbf{Q}}$ to approximate $\mathbf Q$ and run DP to solve the problem approximately. Similarly, we then use the OMP to refine the estimates. The algorithm that runs ${\rm iter}$ iterations of DP and OMP is termed as ``DP+OMP({\rm iter})''.

\begin{figure*}[t!]
		\centering
		\includegraphics[width=0.8\textwidth]{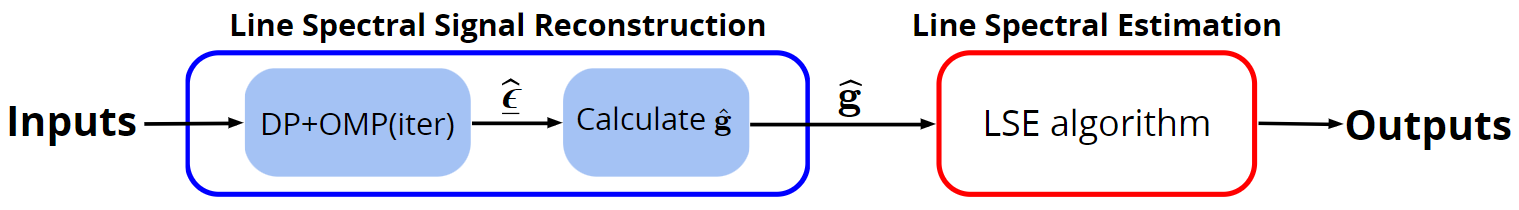}
		\caption{The two-stage USLSE algorithm: The first stage is to perform line spectral signal reconstruction, and the second is to perform LSE.}
		\label{USLSE_figure}
\end{figure*}

Note that DP+OMP(iter) provides the estimate $\widehat{\underline{\boldsymbol{\epsilon}}}$. To recover $\widehat{\boldsymbol{\epsilon}}$ form $\widehat{\underline{\boldsymbol{\epsilon}}}$, the anti-difference operator denoted as $\widehat{{\boldsymbol{\epsilon}}} = S(\widehat{\underline{\boldsymbol{\epsilon}}})$ will be used and is defined as
\begin{subequations}\label{anti_diff}
    \begin{align}
    \widehat{{\epsilon}}_1&= 0\\
    \widehat{{\epsilon}}_{n}&=\sum_{i=1}^{n-1}{\widehat{\underline{\epsilon}}}_i, n = 2,\cdots,N.
\end{align}
\end{subequations}
It should be noted that there is an additive constant ambiguity of $\widehat{\boldsymbol{\epsilon}}$. We can obtain the recovered line spectral signal via $\widehat{\mathbf{g}} = \mathbf{y} + 2\lambda\widehat{\boldsymbol{\epsilon}}$. Finally, a LSE algorithm is applied to obtain $\{\widehat{\boldsymbol{\omega}},\widehat{\mathbf{c}}\}$. Overall, the USLSE algorithm is summarized in Algorithm \ref{USLSE_alg}. USLSE is a two-stage algorithm where line spectral signal reconstruction is performed first, followed by the LSE. The whole process is shown in Fig. \ref{USLSE_figure}.

\begin{algorithm}[ht]
\caption{USLSE Algorithm}\label{USLSE_alg}
\begin{algorithmic}[1]
\STATE \textbf{Inputs:} Modulo samples ${\mathbf y}$, number of sinusoids $K$, and oversampling factor $\gamma$.
\STATE Compute $\underline{\widetilde{\mathbf y}}$.
\STATE Set subset $\mathcal{S}$, state space $\mathcal{V}$, and Markov model order $p$.
\STATE Initialize $\underline{\widehat{\boldsymbol{\epsilon}}} = \mathbf{0}$.
\FOR{${\rm iter}=1:{\rm Iter}_{\rm max}$}
\STATE Compute $\Delta\underline{\widehat{\boldsymbol{\epsilon}}}=$ DP$(\mathbf{z}_{\mathcal{S}} +\mathbf{F}_{\mathcal{S}}\underline{\widehat{\boldsymbol{\epsilon}}},\mathbf{F}_{\mathcal{S}},\mathcal{V},p)$ (Algorithm \ref{DP_alg}).
\STATE Update $\underline{\widehat{\boldsymbol{\epsilon}}} :=\underline{\widehat{\boldsymbol{\epsilon}}}+ \Delta\underline{\widehat{\boldsymbol{\epsilon}}}$.
\STATE Compute $\Delta\underline{\widehat{\boldsymbol{\epsilon}}}$ by solving (\ref{opt_diff_sub_sparse}) (OMP).
\STATE Update $\underline{\widehat{\boldsymbol{\epsilon}}} :=\underline{\widehat{\boldsymbol{\epsilon}}}+ \Delta\underline{\widehat{\boldsymbol{\epsilon}}}$.
\ENDFOR
\STATE Compute ${\widehat{\boldsymbol{\epsilon}}} = S(\underline{\widehat{\boldsymbol{\epsilon}}})$ (up to an additive constant).
\STATE Compute $\widehat{\mathbf{g}} = \mathbf{y} + 2\lambda{\widehat{\boldsymbol{\epsilon}}}$.
\STATE Estimate parameters $\{\widehat{\boldsymbol{\omega}},\widehat{\mathbf{c}}\}$ by applying an LSE algorithm on $\widehat{\mathbf{g}}$.
\STATE \textbf{Outputs:} $\{\widehat{\boldsymbol{\omega}},\widehat{\mathbf{c}}\}$.
\end{algorithmic}
\end{algorithm}

\subsection{Rationality of Approximating ${\mathbf{Q}}$ as $\widetilde{\mathbf{Q}}$}\label{DP_approx_well}
In this subsection, the rationality of approximating ${\mathbf{Q}}$ as $\widetilde{\mathbf{Q}}$ is validated by evaluating the ratio between the squared Frobenius norm of $\widetilde{\mathbf{Q}}_{p}$ to that of $\mathbf{Q}$, where $\widetilde{\mathbf{Q}}_{p}$ is the $p$th order approximation of ${\mathbf{Q}}$ defined in (\ref{Q_tilde}). Proposition \ref{energyratio} provides both the exact expression and a lower bound for the ratio $\frac{\|\widetilde{\mathbf{Q}}_p\|^2_{\rm F}}{\|{\mathbf{Q}}\|^2_{\rm F}}$. Note that for $p=0$, a simple bound is $\frac{\|\widetilde{\mathbf{Q}}_p\|^2_{\rm F}}{\|{\mathbf{Q}}\|^2_{\rm F}}\geq 1-\eta$. In addition, when $N$ and $M$ are significantly larger than $p$, the lower bound can be approximated as a function only with respect to $\eta$ and $p$, i.e., $1-\eta+\frac{8}{\pi^2(1-\eta)}\sum\limits_{k=1}^p\frac{u_{q_k}^2}{k^2}$, based on the fact that $\frac{M-1-k}{N-1-k}\approx \frac{M-1}{N-1} = \eta$ for $k=1,\cdots, p$. For a practical case, where the oversampling factor is $\gamma=10$ and $\eta=\frac{M-1}{N-1}$ is slightly larger than $\frac{1}{\gamma}$ set as $1/10<\eta<1/8$, the results are then listed as follows: for $p=1$, the lower bound is $1-\eta + \frac{8\eta^2}{\pi^2(1 -\eta)}$, and the minimum value is $\approx 0.890 $ with $\eta = 1/8$; for $p=2$, the lower bound is $1-\eta + \frac{16\eta^2}{\pi^2(1 -\eta)}$, and the minimum value is $\approx 0.904 $ with $\eta = 1/8$; for $p=3$, the lower bound is $1-\eta + \frac{24\eta^2}{\pi^2(1 -\eta)}$, and the minimum value is $\approx 0.918 $ with $\eta = 1/8$; for $p=4$, the lower bound is $1-\eta + \frac{32\eta^2}{\pi^2(1 -\eta)}$, and the minimum value is $\approx 0.933 $ with $\eta = 1/8$. It can be seen that the ratio of the squared Frobenius norm of the approximated $\widetilde{\mathbf{Q}}_p$ to that of $\mathbf{Q}$ is at least $0.890$, corresponds to the case when $p=1$. For $p=3$ and $p=4$, the ratio of the squared Frobenius norm of the approximated $\widetilde{\mathbf{Q}}_p$ to that of ${\mathbf{Q}}$ is at least $0.918$. Therefore, approximating ${\mathbf{Q}}$ as $\widetilde{\mathbf{Q}}_p$ is reasonable in this case.

\begin{proposition}\label{energyratio}
Let $M\triangleq N - |\mathcal{S}|$, the ratio of the energy of the banded matrix $\widetilde{\mathbf{Q}}_p$ with bandwidth $p$ to that of $\mathbf Q$ is
\begin{align}\label{Qraequal}
     \frac{\|\widetilde{\mathbf{Q}}_p\|^2_{\rm F}}{\|{\mathbf{Q}}\|^2_{\rm F}} =1-\frac{M-1}{N-1}+\frac{2\sum\limits_{k=1}^{p}\frac{(N-1-k)\sin^2(\frac{k(M-1)\pi}{N-1})}{\sin^2(\frac{k\pi}{N-1})}}{(N-1)^2(N-M)}.
\end{align}
In addition, for $p\leq \frac{N-1}{2}$, $\frac{\|\widetilde{\mathbf{Q}}_p\|^2_{\rm F}}{\|{\mathbf{Q}}\|^2_{\rm F}}$ can be lower bounded by  
\begin{align}\label{pbound}
   \frac{\|\widetilde{\mathbf{Q}}_p\|^2_{\rm F}}{\|{\mathbf{Q}}\|^2_{\rm F}} \geq 1-\eta+\frac{8}{\pi^2}\sum\limits_{k=1}^p\frac{u_{q_k}^2}{k^2(1-\frac{M-1-k}{N-1-k})}
\end{align}
where $\eta\triangleq \frac{M-1}{N-1}$, and 
\begin{align}
    u_{q_k} =
    \begin{cases}
        k\eta - \lfloor k\eta\rfloor,~ &0\leq k\eta - \lfloor k\eta\rfloor < \frac{1}{2}\\
        1 +  \lfloor k \eta\rfloor - k \eta,~ &\frac{1}{2} \leq k\eta - \lfloor k\eta\rfloor < 1
    \end{cases}\notag.
\end{align}
\end{proposition}
\begin{proof}
    The result (\ref{Qraequal}) can be obtained by a straightforward calculation. Based on (\ref{Qraequal}), the lower bound (\ref{pbound}) can be derived as follows
    \begin{align}
           \frac{\|\widetilde{\mathbf{Q}}_p\|^2_{\rm F}}{\|{\mathbf{Q}}\|^2_{\rm F}} &\stackrel{a}\geq 1-\eta+\frac{2}{\pi^2}\sum_{k=1}^p\frac{N-1-k}{N-M}\frac{\sin^2(\frac{k(M-1)\pi}{N-1})}{k^2}\notag\\
           &\stackrel{b}\geq 1-\eta+\frac{8}{\pi^2}\sum\limits_{k=1}^p\frac{u_{q_k}^2}{k^2(1-\frac{M-1-k}{N-1-k})}\notag,
    \end{align}
    where $\stackrel{a}\geq$ is due to $\sin(\frac{k\pi}{N-1})\leq \frac{k\pi}{N-1}$. In addition, $\stackrel{b}\geq$ is derived from the inequality $\frac{2x}{\pi} \leq \sin x, \forall 0 \leq x \leq \frac{\pi}{2}$, and it is established through discussions involving different intervals of $\frac{k(M-1)\pi}{N-1}$.
\end{proof}

\section{Numerical Simulation}
In this section, numerical experiments are conducted to verify the performance of the proposed USLSE. The signal-to-noise ratio (SNR) is defined as $\rm{SNR} = \|\mathbf{x}\|^2_2/\mathbb{E}[\|\mathbf{w}\|^2_2]$, where $\mathbf{x}$ is the noise-free signal and $\mathbf{w}$ is the noise. In this section, we set $\mathbf{w}$ as white Gaussian noise. The normalized mean squared error (NMSE) is $\rm{NMSE}(\widehat{\mathbf{x}}) = \|\widehat{\mathbf{x}}-{\mathbf{x}}\|^2_2/\|{\mathbf{x}}\|^2_2$, where $\widehat{\mathbf{x}}$ denotes the estimate of the algorithm. For the first stage of USLSE, the methods DP, DP+OMP, DP+OMP(iter) are used to solve the optimization problem (\ref{opt_diff_sub}) where ${\rm iter} = 2,3,4$, and the Markov assumption with different orders ($p=1,p=2,p=3,p=4$) are implemented. Given the estimates $\widehat{\underline{\boldsymbol{\epsilon}}}$, the estimates of the simple function $\widehat{\boldsymbol{\epsilon}}$ up to an additive constant can be obtained by using the anti-difference operator (\ref{anti_diff}). The ambiguity is resolved by using $\widehat{\boldsymbol{\epsilon}}:= \widehat{\boldsymbol{\epsilon}} + \lfloor(\sum_{i=1}^N({{\epsilon}}_i - \widehat{{\epsilon}}_i))/N\rceil$, where $ \lfloor\cdot\rceil$ denotes the rounding operator. Finally, a state-of-the-art LSE algorithm will be applied to obtain the estimates $\widehat{\mathbf{x}}$ and $\{\widehat{\boldsymbol{\omega}},\widehat{\mathbf{c}}\}$ based on $\widehat{\mathbf{g}} = \mathbf{y} + 2\lambda\widehat{\boldsymbol{\epsilon}}$. Here, Newtonalized OMP (NOMP) \cite{mamandipoor2016newtonized} due to its fast implementation and high estimation accuracy is used to perform LSE. For different implementations of USLSE, the only difference is the method used to solve the optimization problem (\ref{opt_diff_sub}). Thus, we will use the name of the method solving (\ref{opt_diff_sub}) as the name of different implementations of USLSE. We define the event in which the algorithm successfully recovers the true signal $\mathbf{x}$ provided that $\rm{NMSE}(\widehat{\mathbf{x}})<-15$dB and the probability of successful recovery is adopted as the performance metric\footnote{Different thresholds such as $-10$ dB, $-15$ dB and $-20$ dB are tried in the simulation, and the results are similar.}. The simulation is running on a desktop computer equipped with an 8-core Mac M1 CPU.

Four experiments are conducted to validate the theoretical analysis and verify the excellent performance of USLSE. For the first experiment, the effect of the hyperparameter $\beta$ on USLSE is investigated. Then, the performance of USLSE compared to other state-of-the-art algorithms will be studied. Furthermore, the performance of USLSE versus SNR is investigated in the third experiment. Finally, the reconstruction of bandlimited signals through USLSE is conducted. Parameters of the first and third experiments (Sec. \ref{simu_M} and \ref{versusSNR}) are set as follows: The oversampling factor is $\gamma = 10$, and the number of measurements is $N = 512$. The number of frequencies is $K= 3$, and the frequencies are randomly generated from $(0,2\pi/\gamma)$. The magnitudes of the weight coefficients are drawn i.i.d. from a Gaussian distribution $\mathcal{N}(1, 0.1)$, and the phases are uniformly drawn i.i.d. from $(0, 2\pi)$. The dynamic range of the ADC is set as $\lambda = 0.7$. We set $V = 1$ for USLSE. All results are averaged over $300$ Monte Carlo (MC) trials.

\subsection{The Effect of Hyperparameter $\beta$ on Estimation Performance}\label{simu_M}
\begin{figure*}[!ht]
    \centering
    \includegraphics[width=0.9\textwidth]{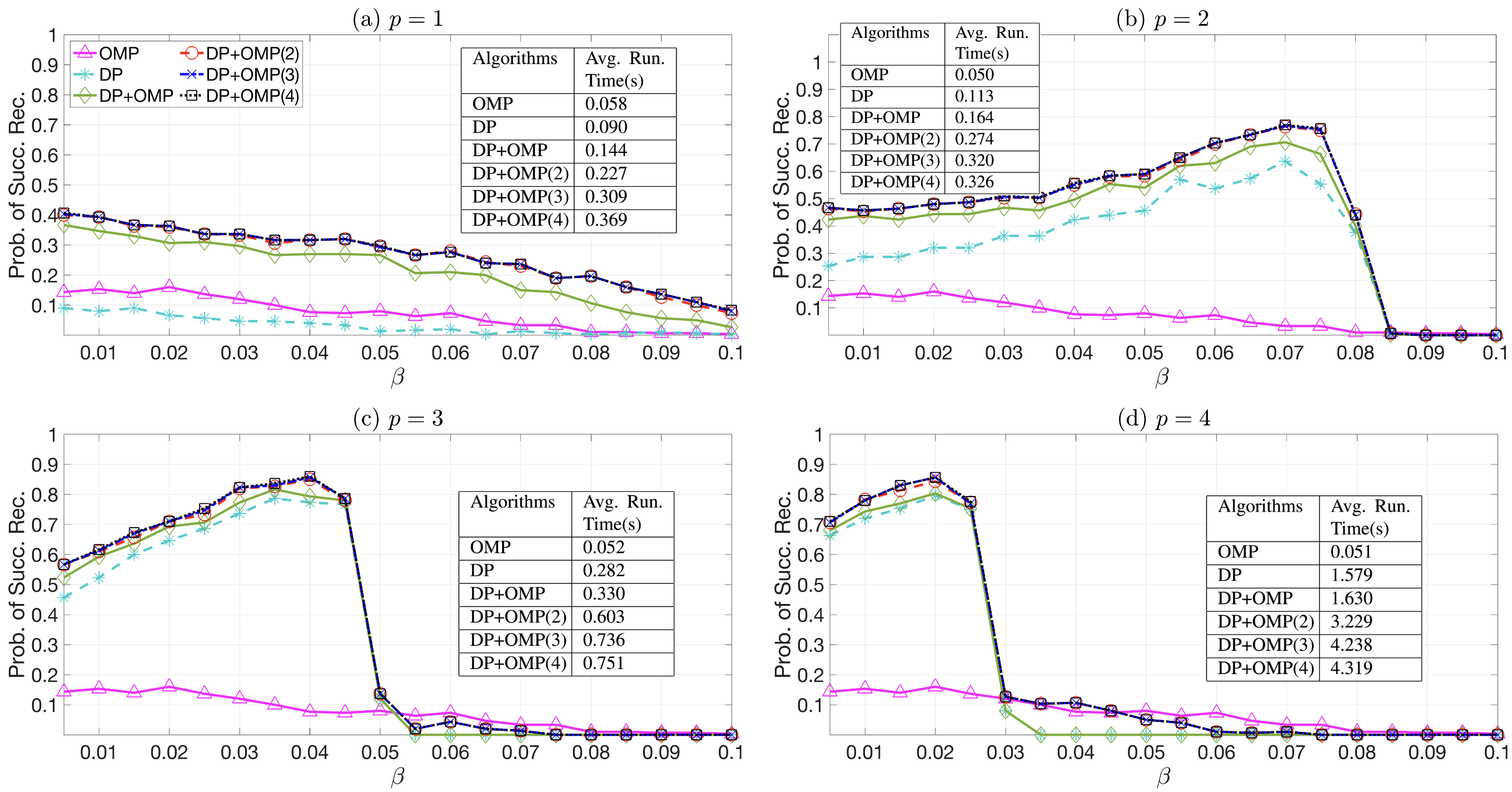}
    \caption{Performance versus $\beta$ for $N = 512$, $\gamma = 10$, $\lambda = 0.7$, and ${\rm SNR} = 22{\rm dB}$. (a)-(d) correspond to the first-order ($p=1$), second-order ($p=2$), third-order ($p=3$), and fourth-order ($p=4$) Markov approximations of $\mathbf{Q}$, respectively.}
    \label{fig_M}
\end{figure*}
According to (\ref{subset_M}), $\frac{1}{N-1}(\approx 0)<\beta<\frac{\gamma-1}{2\gamma}(\approx 0.5)$ is a hyperparameter to set up the optimization problem (\ref{opt_diff_sub}) and should be chosen in advance. We will investigate the performance of the algorithms (OMP, DP, DP+OMP, DP+OMP(iter)) versus $\beta$, where OMP denotes the algorithm that uses the OMP solving problem (\ref{opt_diff_sub}) followed by NOMP to perform the LSE. We set ${\rm SNR} = 22{\rm dB}$, and the probability of successful recovery and the averaged running time are shown in Fig. \ref{fig_M}. To make the subfigures more readable, the result of OMP which is invariant for different $p$ is shown in all the subfigures for comparison.

In Fig. \ref{fig_M}(a) where $p=1$, the probabilities of successful recovery of all algorithms decrease as $\beta$ increases. Besides, DP+OMP(iter) with iter $=2,3,4$ have the same probability of successful recovery and perform best, followed by DP+OMP, OMP, and DP. For $p$ ranging from $2$ to $4$ shown in Fig. \ref{fig_M}(b), Fig. \ref{fig_M}(c) and Fig. \ref{fig_M}(d), respectively, DP+OMP$(2)-(4)$ perform best and achieve similar probability of successful recovery. In detail, the probabilities of successful recovery of DP+OMP$(2)-(4)$ increase first from $0.48$ to $0.78$ as $\beta$ increases to $\beta=0.07$ and then decreases for $p=2$, from $0.58$ to $0.86$ as $\beta$ increases to about $\beta= 0.04$ and then decreases for $p=3$, from $0.7$ to $0.86$ as $\beta$ increases to about $\beta=0.02$ and then decreases for $p=4$. Besides, DP+OMP$(2)-(4)$ benefits from OMP in particular for $p=1$ and $p=2$. The reason is that with a lower value of $p$, the gap between the optimization problem solved by DP (\ref{opt_diff_sub_appro}) and the problem solved by OMP (\ref{opt_diff_sub}) may be larger due to the increased inaccuracy in the approximation of $\mathbf{Q}$. As shown in Fig. \ref{fig_M}, for each $p$, the running time increases in the order of OMP, DP, DP+OMP, DP+OMP(2), DP+OMP(3), DP+OMP(4). Furthermore, the running time of the DP-based algorithms increases as $p$ increases. For all algorithms, the running time of DP+OMP(4) with $p=4$ is longest, which is $4.319s$. 

Except for $p=1$ in Fig. \ref{fig_M}(a), the performances of all DP-based algorithms increase first and then decrease as $\beta$ increases. The reason is that a higher $\beta$ leads to a lower $|\widetilde{\underline{{x}}}_{n}|$ in (\ref{F_delta_g_sub}) and a smaller number of measurements. Thus, there is a trade-off when it comes to selecting the optimal $\beta$ for DP-based algorithms. According to Fig. \ref{fig_M}(b)-(d), as $p$ increases, the optimal $\beta$ decreases, and the optimal probabilities of successful recovery of $p=3$ and $p=4$ are almost the same, which is greater than that of $p=2$. In fact, although as $p$ increases, $\widetilde{\mathbf{Q}}$ approximates $\mathbf{Q}$ better, we can not conclude that a better estimate $\widehat{\underline{\boldsymbol{\epsilon}}}$ will be obtained. We can observe that when $\beta=0.05$, the performance of DP-based algorithms with $p=2$ is better than those with $p=3$, and the performance of DP-based algorithms with $p=3$ is better than those with $p=4$. By carefully investigating the numerical simulation, we found that as $\beta$ increases, DP-based algorithms with larger $p$ tend to overfit and result in poor performance. 

In the following simulations, DP-based algorithms with $p=2, 3, 4$ will be adopted, and the hyperparameter $\beta$ will be set as $\beta = 0.07, 0.04, 0.02$, respectively, corresponding to the optimal choices as shown in Fig. \ref{fig_M}(b)-(d). As the performance of the OMP decreases as $\beta$ increases as shown in Fig. \ref{fig_M}, the set (\ref{subset_M}) of OMP will be set as $\mathcal{S} =\{\lfloor \frac{N-1}{\gamma}\rfloor+2,\cdots,N-1\}$.

\subsection{Comparison with Other Algorithms}
\begin{figure*}[htb!]
    \centering
    \includegraphics[width=\textwidth]{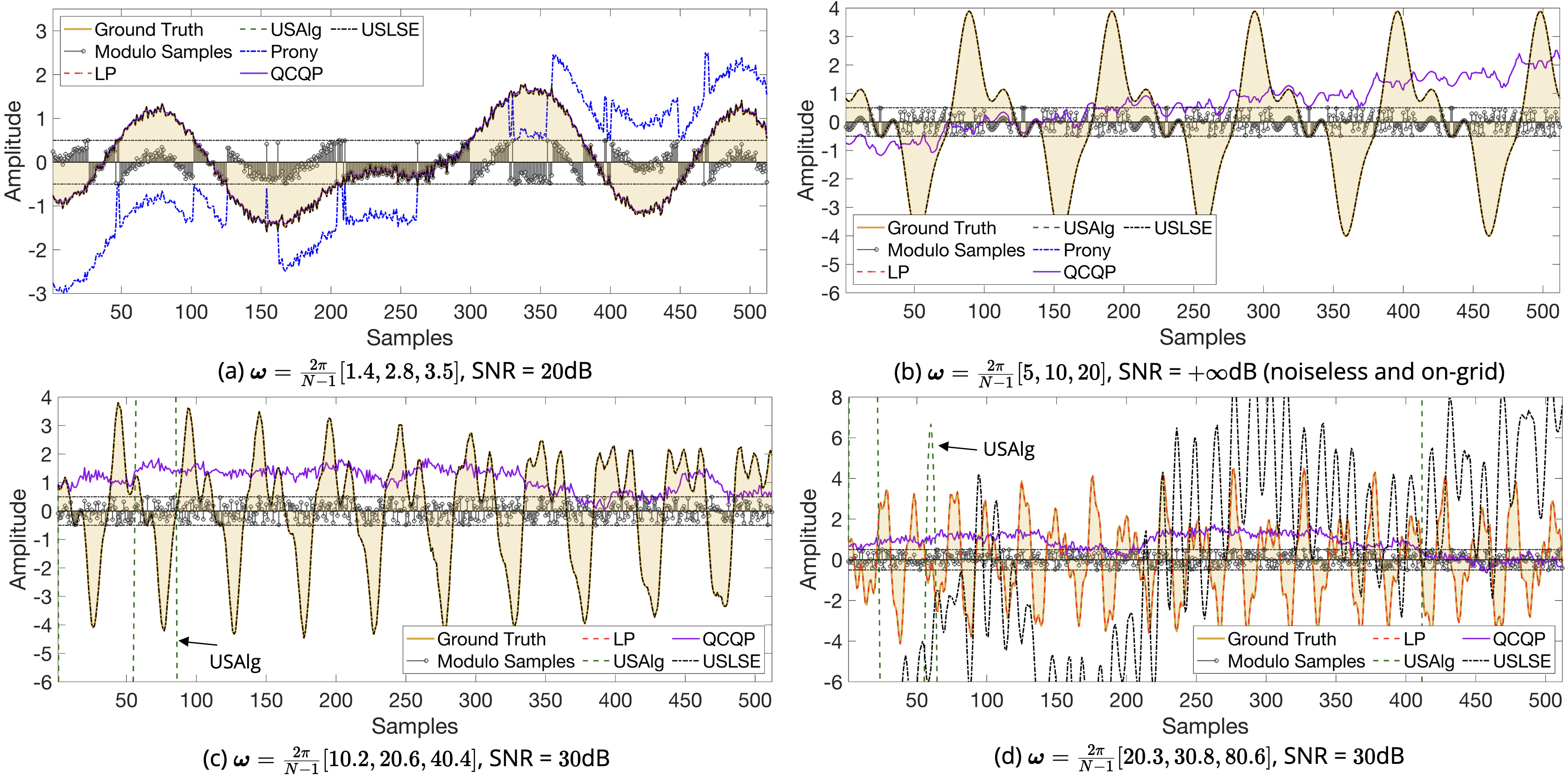}
    \caption{Recovery of line spectral signals (real part) for USLSE and other algorithms. We set $K = 3$, $N=512$, and $\lambda = 0.5$. (a) Noisy and off-grid case with large oversampling factor $\gamma \approx 150$. (b) Noiseless and on-grid case with $\gamma \approx 25$. (c) Noisy and off-grid case with oversampling factor $\gamma \approx 12$. (d) Noisy and off-grid case with a relatively small oversampling factor $\gamma \approx 6$.}
    \label{fig_danci}
\end{figure*}
In this subsection, the USLSE is compared with existing methods such as LP \cite{ordentlich2018modulo}, USAlg \cite{bhandari2020unlimited}, Prony \cite{bhandari2021unlimited}, and QCQP \cite{cucuringu2020provably} that recover the spectrum from modulo samples. It should be noted that $\mathbf{g}$ is a complex vector in our model, while USAlg and QCQP are proposed to deal with the real signal. Thus, we will process the real and imaginary parts of $\mathbf{g}$ separately when using USAlg and QCQP. For the LP method, we assume that the initial few unfolded samples of the signal can be obtained and the autocorrelation function is known. Although the blind version of LP is proposed in \cite{weiss2022blind}, that method requires adaptive adjustment of the ADC's dynamic range, which is not the case in this paper. For USLSE, DP+OMP(2) with $p=3$ is used to solve the problem (\ref{opt_diff_sub}). The recovery results for the real part of the line spectral signal using all algorithms are shown in Fig. \ref{fig_danci}, with similar results observed for the imaginary part.

Fig. \ref{fig_danci}(a) shows the results in a noisy and off-grid scenario with a relatively high oversampling factor $\gamma \approx 150$. LP, USAlg, QCQP, and USLSE successfully recover the original signal, while Prony fails. In Fig. \ref{fig_danci}(b), where the signal is noiseless and the frequencies are precisely on the grid, all algorithms, except QCQP, effectively recover the unfolded signal. As in the cases of Fig. \ref{fig_danci}(c) and Fig. \ref{fig_danci}(d), the cardinality of nonzero elements in $\underline{\boldsymbol{\epsilon}}$, i.e., $\|\underline{\boldsymbol{\epsilon}}\|_0$, is greater than $N/2$, the Prony method can not be utilized in these two scenarios. Fig. \ref{fig_danci}(c) presents the results in a noisy and off-grid setting with $\gamma \approx 12$. LP and USLSE achieve successful recovery of the original signal, while USAlg and QCQP do not. In Fig. \ref{fig_danci}(d) with an SNR of $30$dB and a relatively low oversampling factor $\gamma \approx 6$, all algorithms, except LP, fail to recover the original line spectral signal. In summary, QCQP is effective in recovering signals from noisy modulo samples, but requires a relatively high oversampling factor. Prony's performance is sensitive to noise and relies on sparsity in $\underline{\boldsymbol{\epsilon}}$, leading to its failure in Fig. \ref{fig_danci}(a), (c), and (d). In cases with a large oversampling factor, as in Fig. \ref{fig_danci}(a), or with a noiseless original signal, as in Fig. \ref{fig_danci}(b), USAlg successfully recovers the unfolded samples. However, in Fig. \ref{fig_danci}(c) and Fig. \ref{fig_danci}(d), USAlg fails due to the application of a higher-order differences operator to the noisy original signal. Except for Fig. \ref{fig_danci}(d), where the oversampling factor is relatively small, USLSE effectively recovers the noisy/noiseless signal when frequencies are either on the grid or not on the grid. LP successfully recovers the original signals in all scenarios with a known autocorrelation function. Although USLSE outperforms USAlg, Prony, and QCQP in the settings of Fig. \ref{fig_danci}, it should be noted that USAlg can recover signals with unlimited amplitude when the oversampling factor is larger than $2\pi e$ in the noiseless case, Prony can be used to deal with non-ideal foldings in practical hardware, and QCQP can recover the clean signal from noisy modulo samples.

The running time of all algorithms is also evaluated and the results are shown in Table \ref{Exec_time}. We observe that the execution times for each algorithm remain fairly consistent across various subfigures in Fig. \ref{fig_danci}. Additionally, LP, USAlg, and QCQP exhibit the shortest execution times, followed by Prony and USLSE. Furthermore, the average running times of LP, USAlg, and QCQP are all less than $0.01$s, which are significantly shorter than that of Prony ($=0.233$s) and USLSE ($=0.584$s).
\begin{table}[htb!]
    \begin{center}
\caption{Execution Times of Algorithms in Fig. \ref{fig_danci} (seconds). }\label{Exec_time}
    \begin{tabular}{|c|c|c|c|c|c|c|c|c|}
            \hline
                 \diagbox{Subfig.}{Alg.}& LP & USAlg & Prony & QCQP & USLSE \\ \hline
            Fig. \ref{fig_danci}(a) & 0.004 & 0.007 & 0.125 & 0.003 & 0.574 \\ \hline
            Fig. \ref{fig_danci}(b) & 0.005 & 0.008 & 0.241 & 0.006 & 0.606 \\ \hline
            Fig. \ref{fig_danci}(c) & 0.003 & 0.005 & -- & 0.002 & 0.550 \\ \hline
            Fig. \ref{fig_danci}(d) & 0.005 & 0.006 & -- & 0.003 & 0.604 \\ \hline
             \end{tabular}
    \end{center}
\end{table}

\subsection{Performance Versus SNR}\label{versusSNR}
\begin{figure*}
    \centering
    \includegraphics[width=0.9\textwidth]{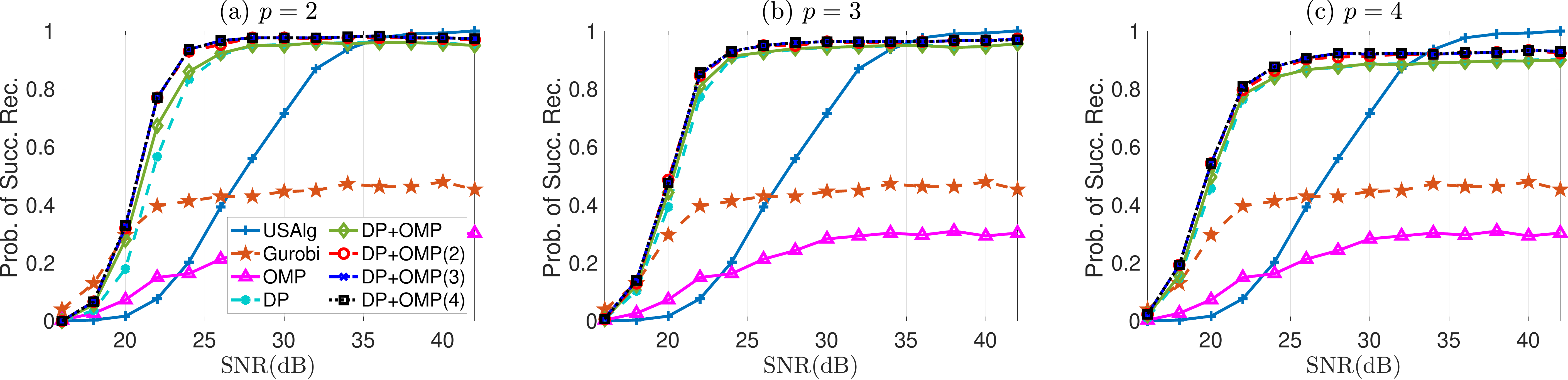}
    \caption{Performance versus SNR for $N = 512$, $\gamma = 10$, and $\lambda = 0.7$. (a)-(c) correspond to the second-order ($p=2$), third-order ($p=3$), and fourth-order ($p=4$) Markov approximations of $\mathbf{Q}$, respectively.}
    \label{fig_SNR}
\end{figure*}

In this subsection, the performances of USLSE under Markov approximation with $p=2,3,4$ versus SNR are investigated, and USAlg will be used as comparison. In addition, problem (\ref{opt_diff_sub}) can be viewed as a mixed integer quadratic programming (MIQP); thus, the professional optimization software Gurobi is also used here to solve it for comparison. The maximum running time of Gurobi is set to be the same as the running time of DP+OMP(4) with $p=4$, which is the longest running time among all other algorithms. To make the subfigures more readable, the results of OMP, Gurobi, and USAlg which are invariant for different $p$ are shown in all the subfigures for comparison. Results are shown in Fig. \ref{fig_SNR}.

Firstly, note that for all subfigures in Fig. \ref{fig_SNR} DP+OMP$(2)-(4)$ are almost identical. In Fig. \ref{fig_SNR}(a), when ${\rm SNR}\leq 20{\rm dB}$, the probabilities of successful recovery of all algorithms are lower than $0.35$. When $22{\rm dB}\leq{\rm SNR}\leq 34{\rm dB}$, DP+OMP$(2)-(4)$ perform best followed by DP+OMP and DP, and all DP-based algorithms outperform USAlg, Gurobi, and OMP. When ${\rm SNR}\geq 36{\rm dB}$, USAlg performs best followed by DP+OMP$(2)-(4)$, DP+OMP, DP, Gurobi and OMP. For Fig. \ref{fig_SNR}(b) and Fig. \ref{fig_SNR}(c), when ${\rm SNR}\leq 18{\rm dB}$, the probabilities of successful recovery of all algorithms are lower than $0.2$. When $20{\rm dB}\leq{\rm SNR}\leq 32{\rm dB}$, DP+OMP$(2)-(4)$ perform best followed by DP+OMP and DP, and all DP-based algorithms outperform USAlg, Gurobi, and OMP in these two subfigures. Furthermore, in Fig. \ref{fig_SNR} (b), USAlg performs best, followed by DP+OMP$(2)-(4)$, DP+OMP, DP, Gurobi, and OMP when ${\rm SNR}\geq 36{\rm dB}$, whereas in Fig. \ref{fig_SNR}(c), similar phenomenon can be observed when ${\rm SNR}\geq 34{\rm dB}$. In summary, the overall performance of Gurobi and OMP is relatively poor, with their respective highest probabilities of successful recovery being $0.48$ and $0.31$. USAlg is sensitive to the noise and has the worst performance when ${\rm SNR}\leq 22{\rm dB}$. However, in the low-noise case, it performs best and achieves the probability of successful recovery equals to $1$ when SNR$=42$dB. For DP-based algorithms, when $22{\rm dB}\leq{\rm SNR}\leq 32{\rm dB}$, they perform better than USAlg, Gurobi, and OMP. It is worth noting that DP-based algorithms cannot achieve a $100\%$ probability of successful recovery for a larger SNR. This is mainly due to two factors. First, the existence of spectral leakage introduces some performance degradation. Second, rather than directly solving the problem (\ref{opt_diff_sub}), we are addressing an approximate version, i.e., problem (\ref{opt_diff_sub_appro}). As SNR increases, the probabilities of successful recovery for DP+OMP$(2)-(4)$ with $p=2$ and $p=3$ gradually converge to approximately $0.97$, while those of DP+OMP$(2)-(4)$ with $p=4$ converge to around $0.93$. Although Proposition \ref{energyratio} suggests that $p=4$ provides a better approximation of $\mathbf{Q}$, its impact on recovery performance remains uncertain. Investigating the influence of the approximation of $\mathbf{Q}$ on signal recovery can be considered for future research.

\subsection{Bandlimited Signals Reconstruction}
\begin{figure*}[htb!]
    \centering
    \includegraphics[width=0.9\textwidth]{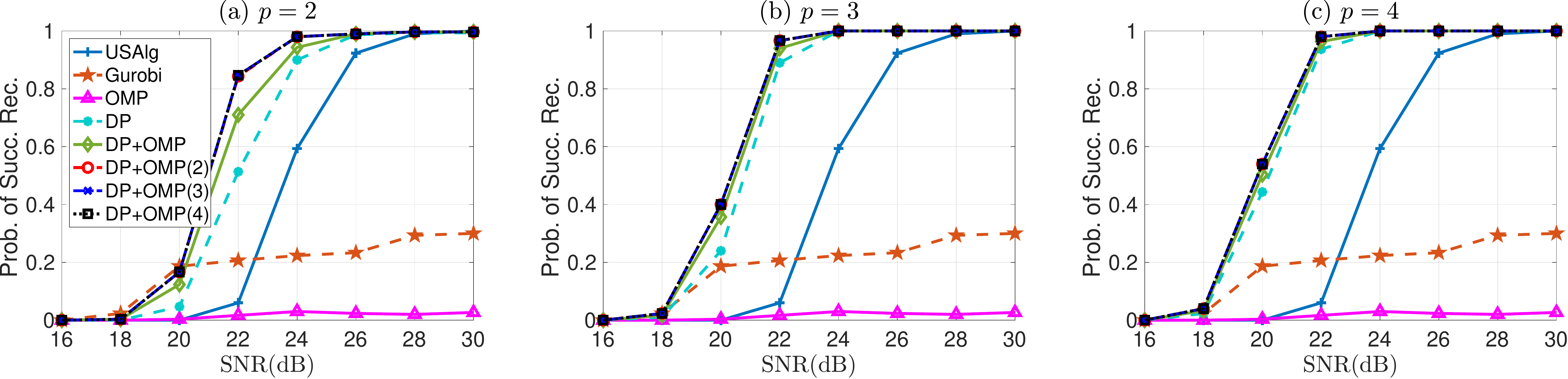}
    \caption{Performance of algorithms for bandlimited signals versus SNR with $N = 400$ and $\gamma = 10$. (a)-(c) correspond to the second-order ($p=2$), third-order ($p=3$), and fourth-order ($p=4$) Markov approximations of $\mathbf{Q}$, respectively.}
    \label{fig_BL}
\end{figure*}
Instead of applying the USLSE on the line spectral signal, our algorithm can also be utilized to recover the bandlimited signal. In this subsection, we will investigate the performances of USAlg, Gurobi, OMP, and all DP-based algorithms under the Markov approximation with $p=2,3,4$ versus SNR on bandlimited signals. The signal is generated as follows. First, we generate a bandlimited real signal $g_{\Omega}(t)$ with bandwidth $\Omega$ \cite{bhandari2021unlimited}. Then, we transform it into the frequency domain $\widetilde{g}_{\Omega}(\omega)$. By setting the amplitude of $\widetilde{g}_{\Omega}(\omega)$ to $0$ for $-\Omega\leq \omega\leq 0$, $\widetilde{g}'_{\Omega}(\omega)$ is obtained. Finally, we obtain the complex bandlimited signal in the time domain ${g}'_{\Omega}(t)$ by applying inverse Fourier transformation on $\widetilde{g}'_{\Omega}(\omega)$. We set $N = 400$ and $\gamma = 10$, and the results are shown in Fig. \ref{fig_BL}. To make the subfigures more readable, the results of OMP, Gurobi, USAlg which are invariant for different $p$ are shown in all the subfigures for comparison.

In Fig. \ref{fig_BL}, OMP can hardly recover the original signal. For all cases, i.e., Fig. \ref{fig_BL}(a)-(c), the probabilities of successful recovery of USAlg, Gurobi, and all DP-based algorithms increase as the SNR increases. When ${\rm SNR}\leq 22{\rm dB}$, DP+OMP$(2)-(4)$ perform best followed by DP+OMP, DP, Gurobi and USAlg for $p=3$ and $p=4$. When ${\rm SNR}\geq 24{\rm dB}$, DP+OMP$(2)-(4)$ perform best followed by DP + OMP, DP, USAlg and Gurobi for $p=2,3,4$. Moreover, when ${\rm SNR}=30{\rm dB}$, all algorithms, except for Gurobi and OMP, achieve perfect successful recovery. Furthermore, the probabilities of the successful recovery of DP-based algorithms increase in our settings as $p$ increases when ${\rm SNR}\leq 24{\rm dB}$. It should be noted that, compared to $p=3$ and $p=4$, DP-based algorithms with $p=2$ benefit more from OMP and iterative refinement.

\section{Empirical Experiments}
Empirical experiments are conducted to demonstrate the performance of USLSE by using an AWR1642 radar. The sampling frequency is $f_s = 10$ MHz. The chirp rate is $\kappa = 30\times 10^{12}$ Hz/s. The maximum radial distance is $r_{\rm{max}} = cf_s/(2\kappa) = 50$m where $c$ is the speed of light. The number of measurements is $N = 256$. Two experiments are conducted. We run the NOMP-CFAR which can maintain the desired false alarm rate without knowledge of the noise variance to process the original measurements and output the range estimates as ground truth \cite{xu2023cfar}. The modulo samples are generated in software from the original measurements.
\subsection{Experiment $1$}
\begin{figure*}[htb!]
    \centering
    \includegraphics[width=0.8\textwidth]{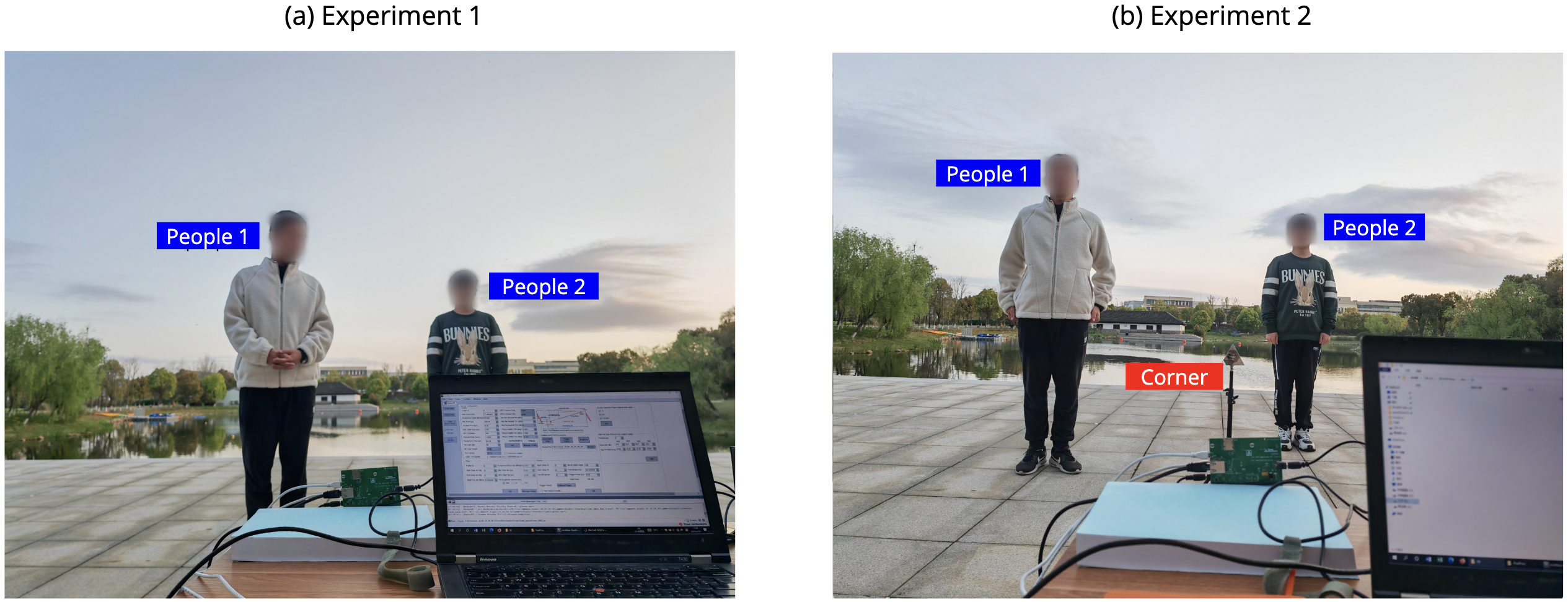}
    \caption{Field experiment setup. (a) There are two targets, referred to as People $1$ and People $2$. (b) There are three targets, referred to as Corner, People $1$, and People $2$. Two people can be regarded as weak signals compared to the corner.}
    \label{fig_Reals}
\end{figure*}
\begin{figure*}[htb!]
    \centering
    \includegraphics[width=\textwidth]{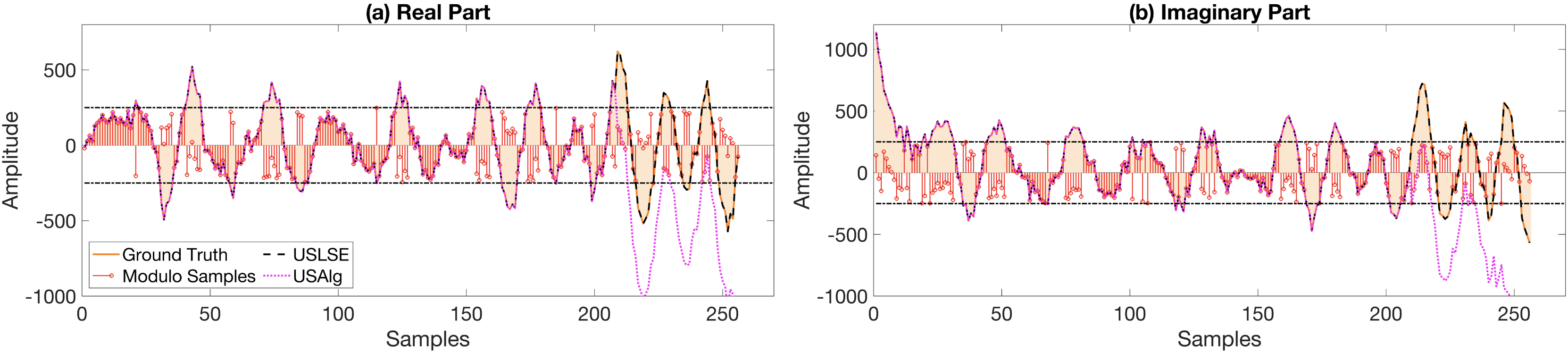}
    \caption{The recovery results of experiment $1$ for the USLSE and USAlg algorithms. For USLSE, DP+OMP(2) with $p=4$ is applied here.}
    \label{fig_Real11}
\end{figure*}
\begin{figure*}[htb!]
\begin{center}
    \centering
    \includegraphics[width=\textwidth]{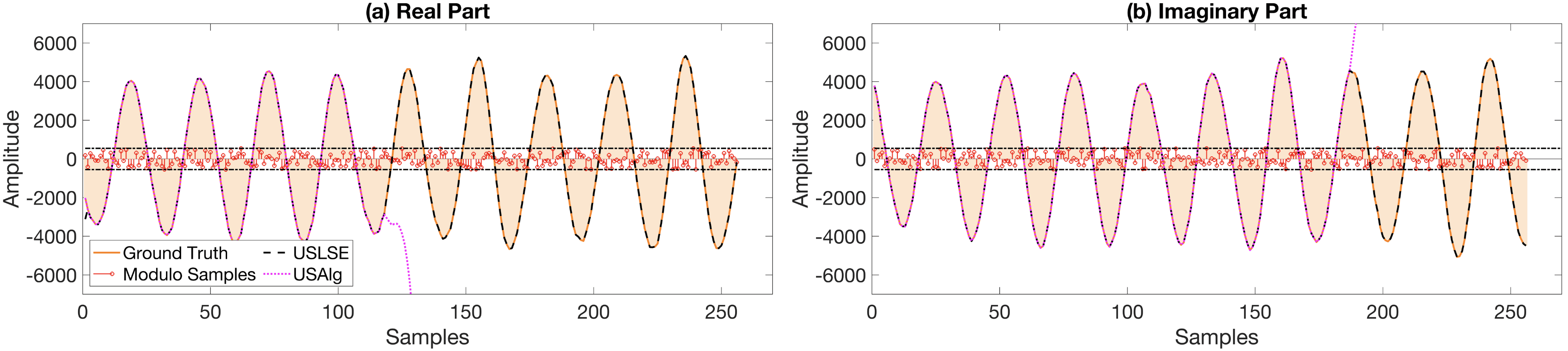}
    \caption{The recovery results of experiment $2$ for the USLSE and USAlg algorithms. For USLSE, DP+OMP(2) with $p=2$ is applied here.}
    \label{fig_Real21}
        \end{center}
\end{figure*}
\begin{figure}[htb!]
    \centering
    \includegraphics[width=0.45\textwidth]{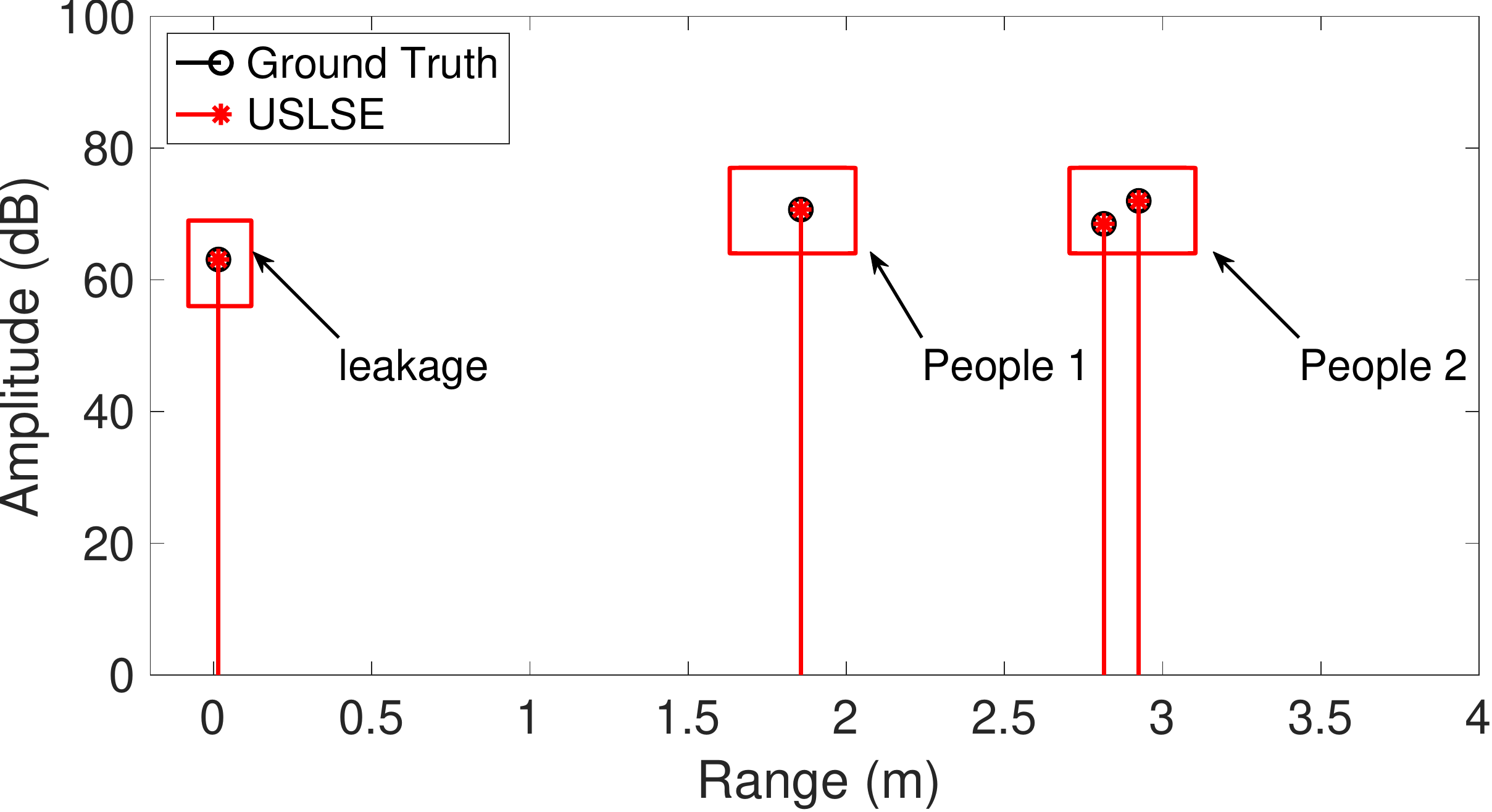}
    \caption{Range estimation results in experiment $1$.}
    \label{fig_Real12}
\end{figure}
The field experiment $1$ is shown in Fig. \ref{fig_Reals}(a), where two people named People $1$ and People $2$ with radial distances being about $1.8$m and $2.8$m are put in front of the radar. The maximum amplitude of the real part and the imaginary part of the real data is about $1200$, and the threshold is set as $\lambda = 250$ which means the maximum folding times is $2$. We set $\gamma = 10$ and $\beta = 0.01$. Reconstruction of signals of USLSE and USAlg algorithms are shown in Fig. \ref{fig_Real11}. For USLSE, we use the fourth-order Markov assumption of $\mathbf{Q}$ and the DP+OMP$(2)$ method is utilized to solve the problem (\ref{opt_diff_sub_appro}). For USAlg, the first-order difference is applied as $D=1$ minimizes both $\|\Delta^D\Re\{\mathbf{g}\}\|_{\infty}$ and $\|\Delta^D\Im\{\mathbf{g}\}\|_{\infty}$ where $\Delta^D\Re\{\mathbf{g}\}$ ($\Delta^D\Im\{\mathbf{g}\}$) is the $D$th-order difference of $\Re\{\mathbf{g}\}$ ($\Im\{\mathbf{g}\}$) defined in \cite{bhandari2020unlimited}. USAlg failed to recover both the real part and the imaginary part, while USLSE successfully recovers both the real and imaginary parts. Besides, we use NOMP-CFAR on the original measurements $\mathbf{g}$ to obtain the ranges of the targets, which are used as the so-called ground truth. In addition, we will also apply NOMP-CFAR on the modulo samples $\mathbf{y}$ and the recovered samples estimated by USLSE and USAlg. The range estimation results ranging from $0$m to $4$m are shown in Fig. \ref{fig_Real12}. For the modulo samples $\mathbf{y}$ and the signal estimated by USAlg, no targets are detected by NOMP-CFAR. For USLSE which perfectly recovers the signal, the estimates are the same as the ground truth. As shown in Fig. \ref{fig_Real12}, the range estimate of People $1$ is about $1.8$m. The ranges of two detected points correspond to People $2$ are $2.8$m and $2.9$m. There also exists a leakage component whose range estimate is $0.01$m.

\subsection{Experiment $2$}
\begin{figure}[htb!]
\begin{center}
    \centering
    \includegraphics[width=0.45\textwidth]{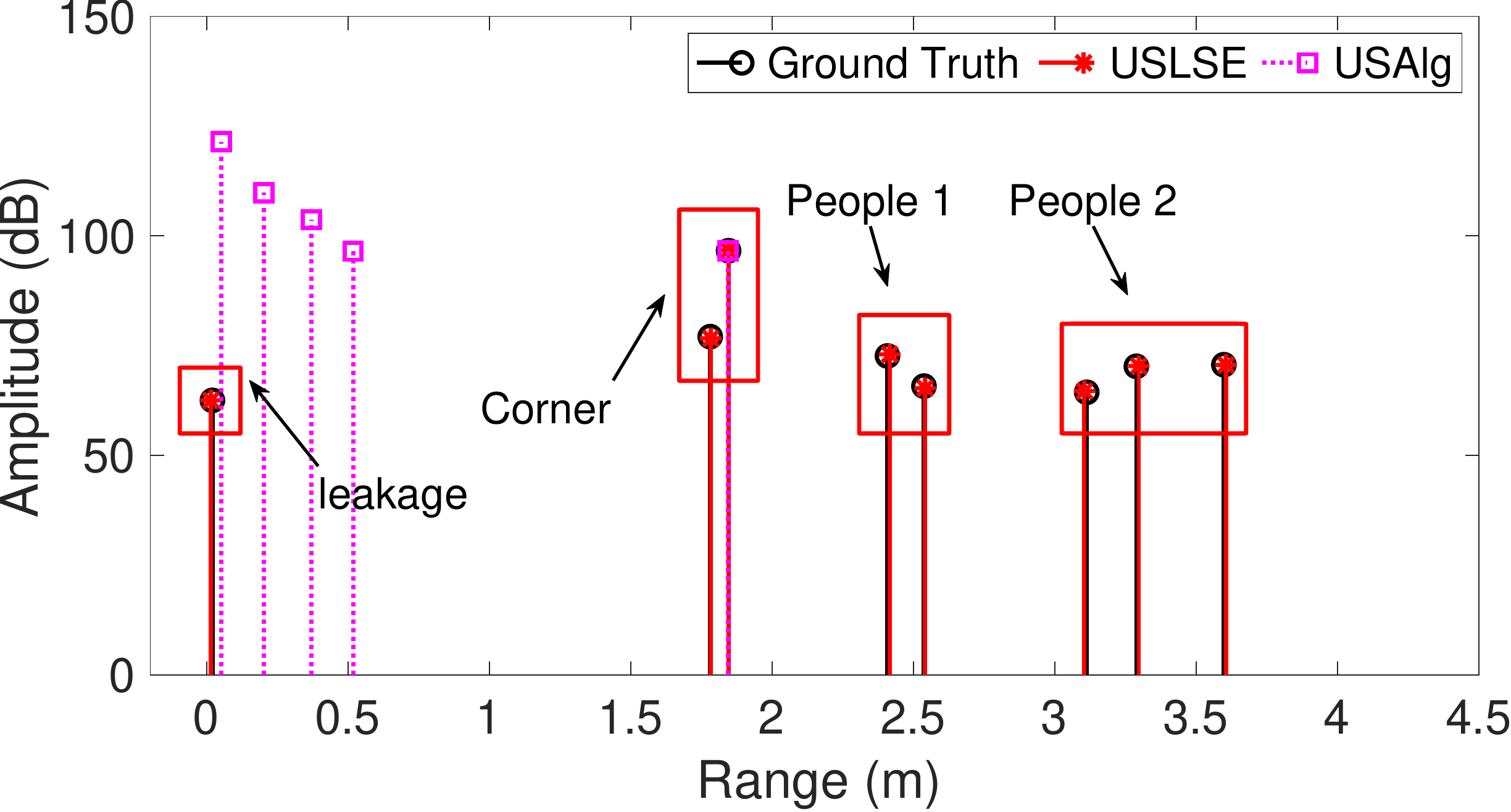}
    \caption{Range estimation results in experiment $2$.}
    \label{fig_Real22}
    \end{center}
\end{figure}
The field experiment consisting of a corner in addition to People 1 and People 2 is shown in Fig. \ref{fig_Reals}(b). The radial distances of Corner, People 1 and People 2 are about $1.8$m, $2.5$m, and $3.3$m, respectively. Compared to the corner, two people can be regarded as weak signals. The maximum amplitude of both the real part and the imaginary part is about $5500$, and the threshold is set as $\lambda = 550$ which means that the maximum folding times is $5$. We set $\gamma = 10$ and $\beta = 0.08$. Reconstruction of signals of USLSE and USAlg algorithms are shown in Fig. \ref{fig_Real21}. For USLSE, we use the second-order Markov assumption of $\mathbf{Q}$ and the DP+OMP$(2)$ method is utilized to solve the problem (\ref{opt_diff_sub_appro}). Similar to experiment $1$, for USAlg, the second-order difference is the optimal choice and is applied here. USAlg failed to recover both the real part and the imaginary part, while USLSE almost successfully recovered the signal except for the small head part of the real part of the signal. The LSE results with ranges ranging from $0$ to $4.5$m are shown in Fig. \ref{fig_Real22}. For the modulo samples $\mathbf{y}$, no targets are detected by NOMP-CFAR. It can be seen that USAlg only detects the strongest corner and generates many false alarms. While USLSE detects Corner, People $1$ and People $2$, and the leakage component. Besides, USLSE does not generate false alarms.
\section{Conclusion}
In this paper, we studied LSE via unlimited sampling. The USLSE algorithm is proposed, which exploits the oversampling and the bandlimited property of the line spectrum to estimate the folding instants and folding times, and then recover the line spectral signal and perform LSE. For USLSE, we novelly combine DP and OMP to iteratively refine the estimates. The robustness of USLSE compared to the existing approach is demonstrated through numerical experiments. Additionally, we also evaluate the performance of USLSE by processing the data acquired by the mmWave radar and show that USLSE is able to jointly estimate two people and one corner with the oversampling factor $10$ and the folding times $5$. 

Several relevant future concerns are worth noting. Firstly, exploring the equivalent conditions for uniquely identifying line spectral signals from modulo samples in future research is worthwhile, as it can serve as a performance benchmark for subsequent algorithms. Secondly, since the algorithm proposed in this paper relies on oversampling, there exists a trade-off between high sampling rates and increased dynamic range. To enable broader applications of our algorithm, a detailed analysis in this regard is worth investigating. Lastly, our proposed algorithm relies on several approximations. Investigating the impact of these approximations on the algorithm's noise resilience and dynamic range enhancement is worth further exploration from an analytical point of view.

\section{Appendix}
\subsection{Proof of Proposition 1 (The Upper Bound of $V$)}\label{state_space_est}
According to the definition of $g[n]$, we have
\begin{align}
    \underline{g}[n] &= g[n+1] - g[n]= \sum_{k=1}^K|c_k|{\rm e}^{{\rm j}\angle c_k}\left({\rm e}^{{\rm j}\omega_k(n+1)} - {\rm e}^{{\rm j}\omega_kn}\right) \notag
\end{align}
for $n= 0,\cdots,N-2$, and the upper bound of $| \underline{g}[n] |$ can be obtained, i.e,
\begin{align}\label{abs_diff_g}
| \underline{g}[n] |=  \left|\sum_{k=1}^K|c_k|{\rm e}^{{\rm j}\angle c_k}{\rm e}^{{\rm j}\omega_kn}\left({\rm e}^{{\rm j}\omega_k} - 1\right)\right|=\left|\sum_{k=1}^K|c_k|{\rm e}^{{\rm j}\angle c_k}{\rm e}^{{\rm j}\omega_kn}{\rm e}^{{\rm j}\omega_k/2}2{\rm j}\sin\left(\frac{\omega_k}{2}\right)\right|\leq Kc_{\rm max}\omega_{\rm max},
\end{align}
where $c_{\rm max}=\underset{k}{\operatorname{max}}|c_k|$ and $\omega_{\rm max} = \underset{k}{\operatorname{max}\omega_k}$. In addition, we have
\begin{align}
|\Re\{\underline{g}[n]\}| \leq Kc_{\rm max}\omega_{\rm max}, |\Im\{\underline{g}[n]\}| \leq Kc_{\rm max}\omega_{\rm max}.
\end{align}
Thus, the upper bound of $V$ can be obtained
\begin{align}
    &V = \max\{\|\Re\{\underline{\boldsymbol{\epsilon}}\}\|_{\infty}, \|\Im\{\underline{\boldsymbol{\epsilon}}\}\|_{\infty}\}  \notag\\
    &\leq \frac{1}{2\lambda}\max\{\|\Re\{\underline{\mathbf{g}}\}\|_{\infty} +\|\Re\{ \underline{\mathbf{y}}\}\|_{\infty}, \|\Im\{\underline{\mathbf{g}}\}\|_{\infty} +\|\Im\{ \underline{\mathbf{y}}\}\|_{\infty}\}  \notag\\
    & \leq \frac{1}{2\lambda}\max\{\|\Re\{\underline{\mathbf{g}}\}\|_{\infty}, \|\Im\{\underline{\mathbf{g}}\}\|_{\infty}\} +1  \notag\\
    & \leq \left\lfloor\frac{K\omega_{\rm max}c_{\rm max}}{2\lambda} + 1\right\rfloor\leq \left\lfloor\frac{K\pi c_{\rm max}}{\lambda\gamma} + 1\right\rfloor.
\end{align}
\subsection{Proof of Proposition 2 (The Upper Bound of $|\widetilde{\underline{x}}[n]|$)}\label{propupperxapp}
Let $\widetilde{\underline{{x}}}[n]$ denote the $(n+1)$th element of $\widetilde{\underline{\mathbf{x}}}$, $|\widetilde{\underline{{x}}}[n]|$ can be upper bounded by
\begin{align}\label{DFT_delta_g}
    &|\widetilde{\underline{{x}}}[n]| = \left|\sum_{m=0}^{N-1}\mathbf{F}_{n+1}\mathbf{J}_{:,m+1}x[m]\right|\notag\\
    &=\frac{1}{\sqrt{N-1}}\left|\sum_{m=0}^{N-2} {\rm e}^{-{\rm j}\frac{2\pi nm}{N-1}}(x[m+1]-x[m])\right|\notag\\
    & \leq  \frac{1}{\sqrt{N-1}}\sum_{k=1}^K\left|\sum_{m=0}^{N-2}c_k {\rm e}^{-{\rm j}\frac{2\pi nm}{N-1}}({\rm e}^{{\rm j}\omega_{k}(m+1)} - {\rm e}^{{\rm j}\omega_{k}m})\right|\notag\\
    & =  \frac{2}{\sqrt{N-1}}\sum_{k=1}^K|c_k|\sin(\frac{\omega_k}{2})\left|\sum_{m=0}^{N-2}{\rm e}^{{\rm j}m(\omega_k - \frac{2\pi n}{N-1})}\right|,
\end{align}
where $n = 0,\cdots, N-2$. Let $\omega_k = \frac{2\pi}{N-1}(a_k+\delta_k)$, where $a_k\in\{0,1,2,\cdots, \lfloor\frac{N-1}{\gamma}\rfloor\}$, and $\delta_k\in[-0.5, 0)\cup(0, 0.5)$, $k=1,\cdots,K$. For $n = 0,\cdots,N-2$, we have
\begin{align}\label{sum_ratio}
    \left|\sum_{m=0}^{N-2}{\rm e}^{{\rm j}m(\omega_k - \frac{2\pi n}{N-1})}\right|= \left|\frac{{1-\rm e}^{{\rm j}(N-1)(\omega_k - \frac{2\pi n}{N-1})}}{1-{\rm e}^{{\rm j}(\omega_k - \frac{2\pi n}{N-1})}}\right|= \frac{|\sin(\pi\delta_k)|}{|\sin(\frac{\pi n}{N-1} - \frac{\omega_k}{2})|}
\end{align}
Inserting (\ref{sum_ratio}) into (\ref{DFT_delta_g}), for $n > \lfloor\frac{N-1}{\gamma}\rfloor$ one obtains
\begin{align}\label{off_tilde_allg}
    |\underline{\widetilde{x}}[n]| \leq&  \frac{2}{\sqrt{N-1}}\sum_{k=1}^K|c_k|\sin(\frac{\omega_k}{2})\frac{|\sin(\pi\delta_k)|}{|\sin(\frac{\pi n}{N-1} - \frac{\omega_k}{2})|}\notag\\
    \stackrel{a}\leq &\frac{2K\pi c_{\rm max}s_{\rm max}\delta_{\rm max}}{\sqrt{N-1}\min\{\sin(\frac{\pi n}{N-1}-\frac{\pi}{\gamma}), \sin(\frac{\pi n}{N-1})\}}
\end{align}
where $\stackrel{a}\leq$ is due to that $\sin(\frac{\pi n}{N-1} - \frac{\omega_k}{2})$ is greater than $0$ and is concave for $\omega_k\in(0,\frac{2\pi}{\gamma})$, $c_{\rm max} \triangleq \operatorname{max}_{i=1}^K|c_i|$, $s_{\rm max}\triangleq \max_{i=1}^K\sin(\frac{\omega_k}{2})$, and $\delta_{\rm max} \triangleq \operatorname{max}_{i=1}^K|\delta_i|$.

\bibliography{USLSEsinglesubmit}

\begin{thebibliography}{10}

\bibitem{stove1992linear}
A.~G. Stove, ``Linear {FMCW} radar techniques,'' {\em IEE Proceedings-F},
  vol.~139, no.~5, pp.~343--350, 1992.

\bibitem{bar2002doa}
O.~Bar-Shalom and A.~J. Weiss, ``{DOA} estimation using one-bit quantized
  measurements,'' {\em IEEE Trans. Aerosp. Electron. Syst.}, vol.~38, no.~3,
  pp.~868--884, 2002.

\bibitem{geroleo2012detection}
F.~G. Geroleo and M.~Brandt-Pearce, ``Detection and estimation of {LFMCW} radar
  signals,'' {\em IEEE Trans. Aerosp. Electron. Syst.}, vol.~48, no.~1,
  pp.~405--418, 2012.

\bibitem{jin2020one}
B.~Jin, J.~Zhu, Q.~Wu, Y.~Zhang, and Z.~Xu, ``One-bit {LFMCW} radar: {S}pectrum
  analysis and target detection,'' {\em IEEE Trans. Aerosp. Electron. Syst.},
  vol.~56, no.~4, pp.~2732--2750, 2020.

\bibitem{zhang2019range}
R.~Zhang, C.~Li, J.~Li, and G.~Wang, ``Range estimation and range-doppler
  imaging using signed measurements in {LFMCW} radar,'' {\em IEEE Trans.
  Aerosp. Electron. Syst.}, vol.~55, no.~6, pp.~3531--3550, 2019.

\bibitem{feuillen2022unlimited}
T.~Feuillen, M.~Alaee-Kerahroodi, A.~Bhandari, B.~Ottersten, {\em et~al.},
  ``Unlimited sampling for {FMCW} radars: {A} proof of concept,'' in {\em Proc.
  IEEE Radar Conf.}, pp.~1--5, 2022.

\bibitem{engels2017advances}
F.~Engels, P.~Heidenreich, A.~M. Zoubir, F.~K. Jondral, and M.~Wintermantel,
  ``Advances in automotive radar: {A} framework on computationally efficient
  high-resolution frequency estimation,'' {\em IEEE Signal Process. Mag.},
  vol.~34, no.~2, pp.~36--46, 2017.

\bibitem{vandersmissen2018indoor}
B.~Vandersmissen, N.~Knudde, A.~Jalalvand, I.~Couckuyt, A.~Bourdoux,
  W.~De~Neve, and T.~Dhaene, ``Indoor person identification using a low-power
  {FMCW} radar,'' {\em IEEE Trans. Geosci. Remote Sens.}, vol.~56, no.~7,
  pp.~3941--3952, 2018.

\bibitem{block2006performance}
F.~J. Block, ``Performance of wideband digital receivers in jamming,'' in {\em
  Proc. IEEE Mil. Commun. Conf.}, pp.~1--7, 2006.

\bibitem{park2007wide}
D.~Park, J.~Rhee, and Y.~Joo, ``A wide dynamic-range {CMOS} image sensor using
  self-reset technique,'' {\em IEEE Electron Device Lett.}, vol.~28, no.~10,
  pp.~890--892, 2007.

\bibitem{yuan2009activity}
J.~Yuan, H.~Y. Chan, S.~W. Fung, and B.~Liu, ``An activity-triggered 95.3 d{B}
  {DR} $-$75.6 d{B} {THD} {CMOS} imaging sensor with digital calibration,''
  {\em IEEE J. Solid-State Circuits}, vol.~44, no.~10, pp.~2834--2843, 2009.

\bibitem{sasagawa2015implantable}
K.~Sasagawa, T.~Yamaguchi, M.~Haruta, Y.~Sunaga, H.~Takehara, H.~Takehara,
  T.~Noda, T.~Tokuda, and J.~Ohta, ``An implantable {CMOS} image sensor with
  self-reset pixels for functional brain imaging,'' {\em IEEE Trans. Electron
  Devices}, vol.~63, no.~1, pp.~215--222, 2015.

\bibitem{krishna2019unlimited}
A.~Krishna, S.~Rudresh, V.~Shaw, H.~R. Sabbella, C.~S. Seelamantula, and C.~S.
  Thakur, ``Unlimited dynamic range analog-to-digital conversion,'' {\em arXiv
  preprint:1911.09371}, 2019.

\bibitem{bhandari2019identifiability}
A.~Bhandari and F.~Krahmer, ``On identifiability in unlimited sampling,'' in
  {\em Proc. Int. Conf. Sampling Theory Appl.}, pp.~1--4, 2019.

\bibitem{romanov2019above}
E.~Romanov and O.~Ordentlich, ``Above the {N}yquist rate, modulo folding does
  not hurt,'' {\em IEEE Signal Process. Lett.}, vol.~26, no.~8, pp.~1167--1171,
  2019.

\bibitem{mulleti2022modulo}
S.~Mulleti and Y.~C. Eldar, ``Modulo sampling of {FRI} signals,'' {\em arXiv
  preprint arXiv:2207.08774}, 2022.

\bibitem{prasanna2020identifiability}
D.~Prasanna, C.~Sriram, and C.~R. Murthy, ``On the identifiability of sparse
  vectors from modulo compressed sensing measurements,'' {\em IEEE Signal
  Process. Lett.}, vol.~28, pp.~131--134, 2020.

\bibitem{bhandari2020unlimited}
A.~Bhandari, F.~Krahmer, and R.~Raskar, ``On unlimited sampling and
  reconstruction,'' {\em IEEE Trans. Signal Process.}, vol.~69, pp.~3827--3839,
  2020.

\bibitem{bhandari2018unlimited1}
A.~Bhandari, F.~Krahmer, and R.~Raskar, ``Unlimited sampling of sparse
  sinusoidal mixtures,'' in {\em Proc. IEEE Int. Symp. Info. Theory},
  pp.~336--340, 2018.

\bibitem{bhandari2018unlimited2}
A.~Bhandari, F.~Krahmer, and R.~Raskar, ``Unlimited sampling of sparse
  signals,'' in {\em Proc. IEEE Int. Conf. Acoust., Speech Signal Process.},
  pp.~4569--4573, 2018.

\bibitem{bouis2021multidimensional}
V.~Bouis, F.~Krahmer, and A.~Bhandari, ``Multidimensional unlimited sampling:
  {A} geometrical perspective,'' in {\em Proc. Eur. Signal Process. Conf.},
  pp.~2314--2318, 2021.

\bibitem{fernandez2021doa}
S.~Fern{\'a}ndez-Menduina, F.~Krahmer, G.~Leus, and A.~Bhandari, ``Do{A}
  estimation via unlimited sensing,'' in {\em Proc. Eur. Signal Process.
  Conf.}, pp.~1866--1870, 2021.

\bibitem{bhandari2021modulo}
A.~Bhandari, M.~Beckmann, and F.~Krahmer, ``The modulo {R}adon transform and
  its inversion,'' in {\em Proc. Eur. Signal Process. Conf.}, pp.~770--774,
  2021.

\bibitem{fernandez2022computational}
S.~Fern{\'a}ndez-Mendui{\~n}a, F.~Krahmer, G.~Leus, and A.~Bhandari,
  ``Computational array signal processing via modulo non-linearities,'' {\em
  IEEE Trans. Signal Process.}, vol.~70, pp.~2168--2179, 2022.

\bibitem{beckmann2020hdr}
M.~Beckmann, F.~Krahmer, and A.~Bhandari, ``{HDR} tomography via modulo {Radon}
  transform,'' in {\em Proc. IEEE Int. Conf. Image Process.}, pp.~3025--3029,
  2020.

\bibitem{bhandari2020hdr}
A.~Bhandari and F.~Krahmer, ``{HDR} imaging from quantization noise,'' in {\em
  Proc. IEEE Int. Conf. Image Process.}, pp.~101--105, 2020.

\bibitem{ordentlich2016integer}
O.~Ordentlich and U.~Erez, ``Integer-forcing source coding,'' {\em IEEE Trans.
  Inf. Theory}, vol.~63, no.~2, pp.~1253--1269, 2016.

\bibitem{ordentlich2018modulo}
O.~Ordentlich, G.~Tabak, P.~K. Hanumolu, A.~C. Singer, and G.~W. Wornell, ``A
  modulo-based architecture for analog-to-digital conversion,'' {\em IEEE J.
  Sel. Topics Signal Process.}, vol.~12, no.~5, pp.~825--840, 2018.

\bibitem{romanov2021blind}
E.~Romanov and O.~Ordentlich, ``Blind unwrapping of modulo reduced {G}aussian
  vectors: {R}ecovering {MSB}s from {LSB}s,'' {\em IEEE Trans. Inf. Theory},
  vol.~67, no.~3, pp.~1897--1919, 2021.

\bibitem{weiss2022blind2}
A.~Weiss, E.~Huang, O.~Ordentlich, and G.~W. Wornell, ``Blind modulo
  analog-to-digital conversion of vector processes,'' in {\em Proc. IEEE Int.
  Conf. Acoust., Speech Signal Process.}, pp.~5617--5621, 2022.

\bibitem{weiss2022blind}
A.~Weiss, E.~Huang, O.~Ordentlich, and G.~W. Wornell, ``Blind modulo
  analog-to-digital conversion,'' {\em IEEE Trans. Signal Process.}, vol.~70,
  pp.~4586--4601, 2022.

\bibitem{bhandari2021unlimited}
A.~Bhandari, F.~Krahmer, and T.~Poskitt, ``Unlimited sampling from theory to
  practice: {F}ourier-{P}rony recovery and prototype {ADC},'' {\em IEEE Trans.
  Signal Process.}, vol.~70, pp.~1131--1141, 2021.

\bibitem{azar2022residual}
E.~Azar, S.~Mulleti, and Y.~C. Eldar, ``Residual recovery algorithm for modulo
  sampling,'' in {\em Proc. IEEE Int. Conf. Acoust., Speech Signal Process.},
  pp.~5722--5726, 2022.

\bibitem{azar2022robust}
E.~Azar, S.~Mulleti, and Y.~C. Eldar, ``Robust unlimited sampling beyond
  modulo,'' {\em arXiv preprint:2206.14656}, 2022.

\bibitem{cucuringu2020provably}
M.~Cucuringu and H.~Tyagi, ``Provably robust estimation of modulo 1 samples of
  a smooth function with applications to phase unwrapping,'' {\em J. Mach.
  Learn. Res.}, vol.~21, no.~32, 2020.

\bibitem{fanuel2022denoising}
M.~Fanuel and H.~Tyagi, ``Denoising modulo samples: k-{NN} regression and
  tightness of {SDP} relaxation,'' {\em Information and Inference: A Journal of
  the IMA}, vol.~11, no.~2, pp.~637--677, 2022.

\bibitem{mulleti2023hardware}
S.~Mulleti, E.~Reznitskiy, S.~Savariego, M.~Namer, N.~Glazer, and Y.~C. Eldar,
  ``A hardware prototype of wideband high-dynamic range analog-to-digital
  converter,'' {\em IET Circuits, Devices Syst.}, vol.~17, no.~4, pp.~181--192,
  2023.

\bibitem{florescu2022surprising}
D.~Florescu, F.~Krahmer, and A.~Bhandari, ``The surprising benefits of
  hysteresis in unlimited sampling: {T}heory, algorithms and experiments,''
  {\em IEEE Trans. Signal Process.}, vol.~70, pp.~616--630, 2022.

\bibitem{florescu2022unlimited1}
D.~Florescu and A.~Bhandari, ``Unlimited sampling with local averages,'' in
  {\em Proc. IEEE Int. Conf. Acoust., Speech Signal Process.}, pp.~5742--5746,
  2022.

\bibitem{florescu2022unlimited2}
D.~Florescu and A.~Bhandari, ``Unlimited sampling via generalized
  thresholding,'' in {\em Proc. IEEE Int. Symp. Inf. Theory}, pp.~1606--1611,
  2022.

\bibitem{shtendel2023unlimited}
G.~Shtendel, D.~Florescu, and A.~Bhandari, ``Unlimited sampling of bandpass
  signals: {Computational} demodulation via undersampling,'' {\em IEEE Trans.
  Signal Process.}, vol.~71, pp.~4134--4145, 2023.

\bibitem{mamandipoor2016newtonized}
B.~Mamandipoor, D.~Ramasamy, and U.~Madhow, ``Newtonized orthogonal matching
  pursuit: {F}requency estimation over the continuum,'' {\em IEEE Trans. Signal
  Process.}, vol.~64, no.~19, pp.~5066--5081, 2016.

\bibitem{xu2023cfar}
M.~Xu, J.~Zhu, J.~Fang, N.~Zhang, and Z.~Xu, ``{CFAR} based {NOMP} for line
  spectral estimation and detection,'' {\em IEEE Trans. Aerosp. Electron.
  Syst.}, vol.~59, no.~5, pp.~6971--6990, 2023.

\end{thebibliography}
\bibliographystyle{ieeetr}
\end{document}